\documentclass[letterpaper,11pt]{article}
\usepackage{amssymb}
\usepackage{amsmath}
\usepackage{amstext}
\usepackage{graphicx,epsfig}
\usepackage{epsfig}
\usepackage{verbatim} 
\usepackage{fancybox}
\usepackage{color}
\usepackage{ulem}
\usepackage{enumitem}
\usepackage{subfigure}
\usepackage{bbm}
\usepackage{parskip}
\usepackage{dsfont}
\usepackage[numbers,sort&compress]{natbib}
\usepackage[body={6.5in, 9in}, right=1in, top=1in]{geometry}

\newcommand{\be}{\begin{equation}}
\newcommand{\ee}{\end{equation}}
\newcommand{\bea}{\begin{eqnarray}}
\newcommand{\eea}{\end{eqnarray}}
\newcommand{\beas}{\begin{eqnarray*}}
\newcommand{\eeas}{\end{eqnarray*}}

\def\({\left(}
\def\){\right)}

\def\gsim{ \lower .75ex \hbox{$\sim$} \llap{\raise .27ex \hbox{$>$}} }
\def\lsim{ \lower .75ex \hbox{$\sim$} \llap{\raise .27ex \hbox{$<$}} }

\usepackage{xcolor,colortbl}

\begin{document}
\def\thefootnote{\fnsymbol{footnote}}

\begin{center}
\Large{\textbf{Theory of Dark Matter Superfluidity}} \\[0.5cm]
 
\large{Lasha Berezhiani and Justin Khoury}
\\[0.5cm]

\small{
\textit{Center for Particle Cosmology, Department of Physics and Astronomy, \\ University of Pennsylvania, Philadelphia, PA 19104}}

\vspace{.2cm}

\end{center}

\vspace{.2cm}

\hrule \vspace{0.2cm}
\centerline{\small{\bf Abstract}}
{\small We propose a novel theory of dark matter (DM) superfluidity that matches the successes of the $\Lambda$CDM model on cosmological scales while simultaneously reproducing the MOdified Newtonian Dynamics (MOND) phenomenology on galactic scales. The DM and MOND components have a common origin, representing different phases of a single underlying substance. DM consists of axion-like particles
with mass of order eV and strong self-interactions. The condensate has a polytropic equation of state $P\sim \rho^3$ giving rise to a superfluid core within galaxies. Instead of behaving as individual collisionless
particles, the DM superfluid is more aptly described as collective excitations. Superfluid phonons, in particular, are assumed to be governed by a MOND-like effective action and mediate a MONDian acceleration between
baryonic matter particles. Our framework naturally distinguishes between galaxies (where MOND is successful) and galaxy clusters (where MOND is not): due to the higher velocity dispersion in clusters, and correspondingly higher temperature, the DM in clusters is either in a mixture of superfluid and normal phase, or fully in the normal phase. The rich and well-studied physics of superfluidity leads to a number of observational signatures: array of low-density vortices in galaxies, merger dynamics that depend on the infall velocity vs phonon sound speed; distinct mass peaks in bullet-like cluster mergers, corresponding to superfluid and normal components; interference patters in super-critical mergers. Remarkably, the superfluid phonon effective theory is strikingly similar to that of the unitary Fermi gas, which has attracted much excitement in the cold atom community in recent years. The critical temperature for DM superfluidity is of order mK, comparable to known cold atom Bose-Einstein condensates. Identifying a precise cold atom analogue would give important insights on the microphysical interactions underlying DM superfluidity. Tantalizingly, it might open the possibility of simulating the properties and dynamics of galaxies in laboratory experiments.}
\vspace{0.3cm}
\noindent
\hrule
\def\thefootnote{\arabic{footnote}}
\setcounter{footnote}{0}

\section{Introduction}

The most clear-cut evidence for dark matter (DM) comes from observations on the largest scales. The standard $\Lambda$-Cold-Dark-Matter ($\Lambda$CDM) model, in which DM consists of collisionless particles, does exquisitely well at fitting the background expansion history, the detailed shape of microwave background and matter power spectra, as well as the abundance and mass function of galaxy clusters. On smaller scales, however, the situation is murkier. As simulations and observations of galaxies have improved, a number of challenges have emerged for the CDM paradigm. 

\begin{figure}[t]
\centering
\includegraphics[width=3.5in]{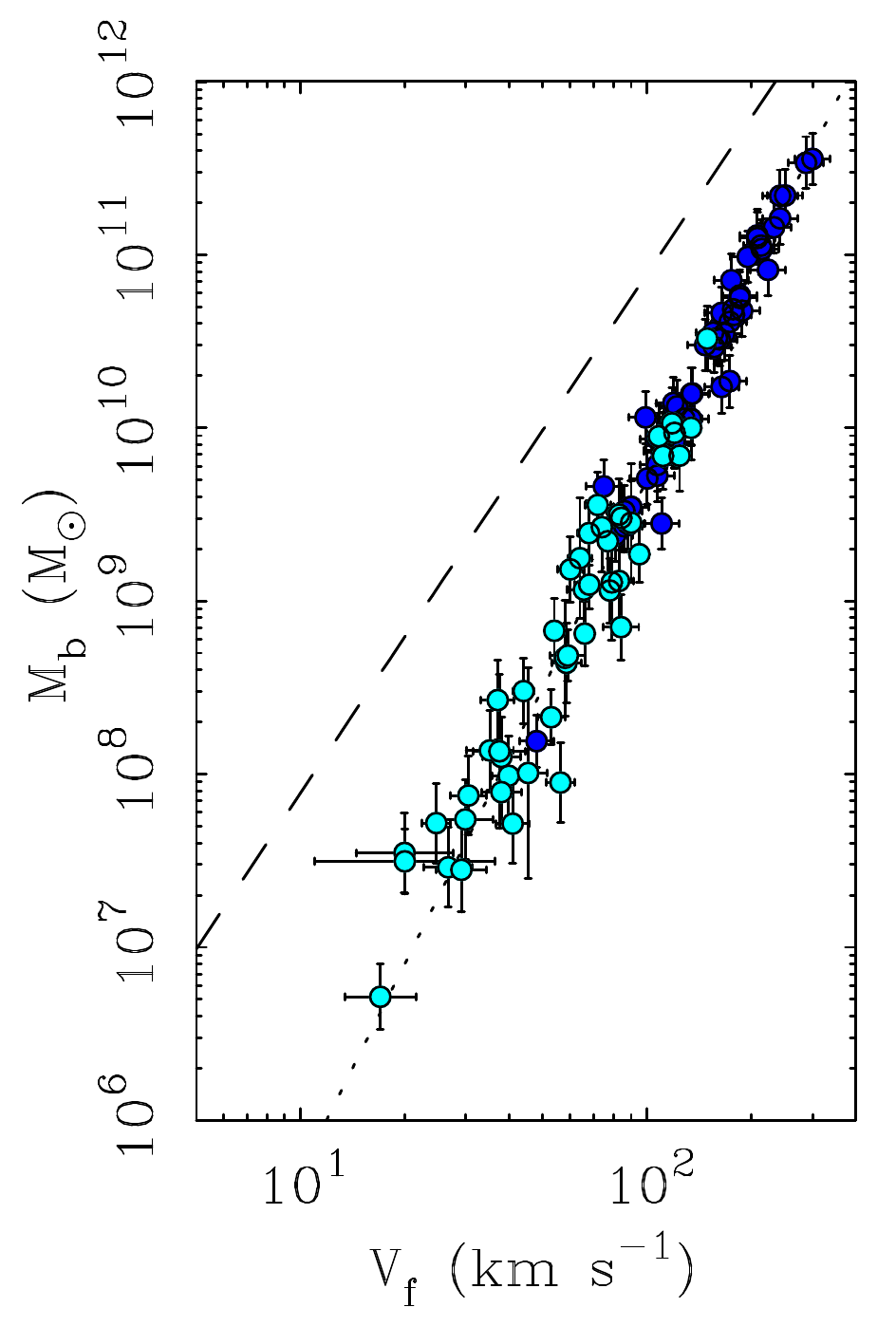}
\caption{\label{BTFRdSph} \small The Baryonic-Tully-Fisher-Relation (BTFR), reproduced from~\cite{Famaey:2011kh}. The dark blue points are star-dominated galaxies; the light blue circles are gas-dominated. The dashed line has a slope of 3, corresponding to the $\Lambda$CDM prediction. The dotted line has slope 4, in good agreement with the data.}
\end{figure}

For starters, galaxies in our universe are observed to be remarkably regular, a fact embodied by various empirical scaling relations. The most striking example is the Baryonic Tully Fisher Relation (BTFR)~\cite{Freeman1999,McGaugh:2000sr,McGaugh:2005qe,McGaugh:2011ac}, which relates the baryonic mass $M_{\rm b}$ to the asymptotic circular velocity $v_{\rm c}$:\footnote{The BTFR extends the old Tully-Fisher relation~\cite{Tully:1977fu}, relating the optical luminosity to velocity as $L\sim v^4_{\rm c}$. Since the mass-to-light ratio is not constant among different types of galaxies, the inferred slope and scatter end up depending on the choice of band filter. Replacing luminosity by total baryonic mass~\cite{Freeman1999,McGaugh:2000sr} ({\it i.e.}, stars and gas) reduces the scatter and extends the validity the scaling relation over many decades in mass~\cite{McGaugh:2011ac}, as shown in Fig.~\ref{BTFRdSph}.}
\be
M_{\rm b} \sim v^4_{\rm c}\,.
\label{btf}
\ee
Figure~\ref{BTFRdSph}, reproduced from~\cite{Famaey:2011kh}, shows excellent agreement with remarkably little scatter in the high-mass end comprised of star-dominated (dark blue circles) and gas-dominated disc galaxies (light blue circles). On the theory side, the standard collapse model predicts a scaling between the total mass (dark plus baryonic) and circular velocity at the virial radius: $M_{\rm vir} \sim v_{\rm vir}^3$. Despite the different slope, this is not {\it a priori} inconsistent with~\eqref{btf} since the translation from virial parameters to observables can be mass-dependent. However, the real challenge for $\Lambda$CDM lies in explaining the remarkably small level of scatter around this slope in the high-mass end, as shown in Fig.~\ref{BTFRdSph}. How can baryonic feedback processes, which are inherently stochastic, result in such a tight correlation across different galaxy types? Indeed, recent hydrodynamical simulations~\cite{Vogelsberger:2014dza} show considerably larger scatter than observations~\cite{McGaugh:2011ac}.

Another set of challenges comes from dwarf satellite galaxies in the Local Group. Dwarf satellites are highly DM-dominated objects and thus well-suited to detailed tests of DM microphysics. As the old ``missing satellite" problem~\cite{Kauffmann:1993gv,Klypin:1999uc,Moore:1999nt} has gradually been alleviated through the discovery of ultra-faint dwarfs~\cite{Willman:2004kk,Belokurov:2006ph,Tollerud:2008ze,Walsh:2008qn}, new sharper problems have emerged. Recent attempts at matching the populations of simulated subhaloes and observed MW dwarf galaxies have revealed a ``too big to fail'' problem~\cite{BoylanKolchin:2011de,toobigtofail2}: the most massive dark halos are too dense to host the brightest MW satellites. Even more puzzling is the fact that the majority of the MW~\cite{Kroupa:2004pt,Pawlowski:2012vz,Pawlowski:2013kpa,Pawlowski:2013cae} and Andromeda (M31)~\cite{Ibata:2013rh,Conn:2013iu,Ibata:2014pja} satellites lie within vast planar structures and are co-rotating within these planes.\footnote{Phase-space correlated dwarfs have also been found around galaxies beyond the Local Group~\cite{Ibata:2014csa}.} This is puzzling for $\Lambda$CDM, though mechanisms have been proposed~\cite{Libeskind:2005hs,Zentner:2005wh,Libeskind:2010pd,Lovell:2010ap}.

Another puzzle comes from tidal dwarfs --- ``recycled" galaxies that form in the tidal material created by merging spirals.
While standard theory tells us that tidal dwarfs should be devoid of dark matter~\cite{Barnes,Duc:2004kd,Bournaud:2006qz}, recent observations of three such objects around NGC5291~\cite{Bournaud:2007sz} have revealed a dynamical mass discrepancy of about 2-3 times the visible mass. A standard explanation is that the missing matter is in the form of cold baryonic gas~\cite{Bournaud:2007sz}, however this seems unlikely given their flat rotation curves and the remarkable consistency with the BTFR~\cite{Gentile:2007gp}. One should be wary of drawing definitive conclusions from a few objects, but a larger sample of tidal dwarfs will reduce uncertainties and can provide a critical test of $\Lambda$CDM. 

\subsection{MOND: successes, challenges and failures}

A radical alternative is MOdified Newtonian Dynamics (MOND)~\cite{Milgrom:1983ca,Milgrom:1983pn,Milgrom:1983zz,Bekenstein:1984tv}, which proposes to replace DM with a modification of the Newtonian force law. The force law is standard at large acceleration ($a\simeq a_{\rm N}$ for $a_{\rm N} \gg a_0$) but modified at low acceleration ($a\simeq \sqrt{a_{\rm N}a_0}$ for $a_{\rm N} \ll a_0$). This empirical force law has been remarkably successful at explaining a wide range of galactic phenomena~\cite{Sanders:2002pf,Famaey:2011kh}. For spiral galaxies, it predicts asymptotically flat rotation curves and 
provides an excellent fit to detailed rotation curves~\cite{Sanders:2002pf}. The critical acceleration $a_0$ is the only free parameter (apart from the ${\cal O}(1)$ mass-to-light ratio for each galaxy), with the best-fit value intriguingly of order the present Hubble parameter:
\be
a_0 \simeq \frac{1}{6}H_0 \simeq  1.2\times 10^{-8}~{\rm cm}/{\rm s}^2\,.
\label{critacc}
\ee
The BTFR is an exact consequence of this force law --- deep in the MOND regime ($a_{\rm N} \ll a_0$), a test particle will orbit an isolated spherically-symmetric source according to $\frac{v^2_{\rm c}}{r} = \sqrt{\frac{G_{\rm N} M_{\rm b} a_0}{r^2}}$, and hence
\be
M_{\rm b} = \frac{v^4_{\rm c}}{G_{\rm N} a_0}\,.
\label{MONDbtf}
\ee

The vast planar structures seen around the MW and Andromeda also find a plausible explanation in MOND, as the result of tidal stripping during a fly-by encounter between these galaxies. With the MOND force law, this encounter has been estimated to have occurred $\sim 10$~Gyr ago, with $\lsim \; 55$~kpc closest approach distance~\cite{Zhao:2013uya}. Unlike in $\Lambda$CDM, where galaxies are surrounded by extended DM halos and dynamical friction would cause a rapid merger, in MOND there is only stellar dynamical friction and a merger can be avoided~\cite{Nipoti:2007ik,Tiret:2007fy,Combes:2009ab}. MOND predicts that tidal dwarf galaxies should have flat rotation curves and fall on the BTFR, consistent with the NGC5291 dwarfs~\cite{Bournaud:2007sz,Gentile:2007gp}. 

On the flip side, dwarf satellites, particularly the MW dwarf spheroidals, have long posed a challenge for MOND~\cite{Spergel,Milgrom:1995hz,Angus:2008vs,Hernandez:2009by,Lughausen:2014hxa}. Five of the classical dwarfs are consistent with the BTFR, but two (Draco and Ursa Minor) fall below it~\cite{Spergel,Milgrom:1995hz}. Nearly all the ultra-faint dwarfs lie systematically below the BTFR~\cite{McGaugh:2010yr}. However, the derivation of the BTFR in MOND assumes dynamical equilibrium, whereas the discrepant dwarfs may be undergoing tidal disruption~\cite{McGaugh:2010yr}. Moreover, velocity estimates for these objects are complicated by interlopers~\cite{Serra:2009tj}. On the other hand, MOND does an excellent job at explaining the observed velocity dispersions in Andromeda's dwarf satellites~\cite{McGaugh:2013wj,McGaugh:2013zqa}. Finally, globular clusters also pose a challenge for MOND~\cite{Ibata:2011ri}. 

MOND faces much more severe challenges on extra-galactic scales. To reproduce the observed temperature profile of galaxy clusters~\cite{Aguirre:2001fj}, one must invoke some form of dark matter, either as massive neutrinos~\cite{Sanders:2002ue,Angus:2006ev,Angus:2011hx} and/or cold dense gas clouds~\cite{Milgrom:2007cs}. Relativistic versions of MOND, such as the Tensor-Vector-Scalar (TeVeS) theory~\cite{Sanders:1996wk,Bekenstein:2004ne,Moffat:2005si,Sanders:2005vd,Zlosnik:2006sb,Skordis:2008pq,Contaldi:2008iw} and other related proposals~\cite{Milgrom:2009gv,Blanchet:2009zu,Deffayet:2011sk,Blanchet:2011wv} (see ~\cite{Bruneton:2007si} for a review), cannot match the CMB power spectrum~\cite{Skordis:2005xk,Zuntz:2010jp}. Without a significant dark matter component, the baryonic oscillations in the matter power spectrum tend to be far too pronounced~\cite{Skordis:2005xk,Dodelson:2011qv}. Finally, numerical simulations of MONDian gravity with massive neutrinos fail to reproduce the observed cluster mass function~\cite{Angus:2013sxa,Angus:2014kja}.

\subsection{DM-MOND hybrids}
\label{best}

What we have learned is that MOND and CDM are each successful in almost mutually exclusive regimes. The $\Lambda$CDM model successfully explains the expansion and linear growth histories, as well as the abundance of clusters, but faces a number of challenges on galactic scales. MOND does very well overall at explaining the observed properties of galaxies, in particular the empirical scaling relations, but  it seems highly improbable that it can ever be made consistent with the detailed shape of the
CMB and matter power spectra.

This has led various people to propose hybrid models that include both DM {\it and} MOND phenomena~\cite{Blanchet:2006yt,Blanchet:2008fj,Zhao:2008rq,Bruneton:2008fk,Li:2009zzh,Ho:2010ca,Ho:2011xc,Ho:2012ar}. For instance, one of us recently proposed such a hybrid model, involving two scalar fields~\cite{Khoury:2014tka}: one scalar field acts as DM, the other mediates a MOND-like force law. This model enjoys a number of advantages compared to TeVeS and other relativistic MOND theories. For starters, it only requires two scalar fields, as opposed to the scalar and vector fields of TeVeS. Secondly, unlike TeVeS, its predictions on cosmological scales are consistent with observations, thanks to the DM scalar field. Finally, the model offers a better fit to the temperature profile of galaxy clusters.

The improved consistency with data does come at the price of having two {\it a priori} distinct components --- a DM-like component and a modified-gravity component. It would be much more compelling if these two components somehow had a common origin. Furthermore, the theory must be adjusted such as to avoid co-existence of DM-like and MOND-like behavior. This requires that the parameters of the theory be mildly scale or mass dependent, which adds another layer of complexity. 

\subsection{Unified approach: MOND phenomenon from DM superfluidity}

In this paper, along with its shorter companion~\cite{Berezhiani:2015pia}, we propose a unified framework for the DM and MOND phenomena. The DM and MOND components have a common origin, representing different phases of a single underlying substance. This is achieved through the rich and well-studied physics of superfluidity. 

There are two central ideas underlying this work. The first idea is quite general, namely that DM forms a superfluid inside galaxies
with a coherence length of order the size of galaxies. As we will see, the phenomenon of DM superfluidity is quite generic if the DM particle is sufficiently light and has
sufficiently strong self-interaction. Specifically, as a back-of-the-envelope calculation, we can estimate the condition for the onset of superfluidity by ignoring interactions among DM particles. With this simplifying approximation, the requirement for superfluidity amounts to demanding that the de Broglie wavelength $\lambda_{\rm dB} \sim \frac{1}{mv}$ of DM particles should overlap. Using the typical velocity $v$ and density of DM particles in galaxies, this translates into an upper bound $m \;\lsim\; 2~{\rm eV}$ on the DM particle mass. 

Another requirement for Bose-Einstein condensate is that DM thermalize within galaxies. We assume that DM particles interact through contact repulsive interactions.
Demanding that the interaction rate be larger than the galactic dynamical time places a lower bound of $\frac{\sigma}{m} \gsim\; 0.1~{\rm cm}^2/{\rm g}$. This is just
below the most recent constraint $\lsim\; 0.5~{\rm cm}^2/{\rm g}$ from galaxy cluster mergers~\cite{Harvey:2015hha}, though we will argue such
constraints must be carefully reanalyzed in the superfluid context. 

Again ignoring interactions, the critical temperature for DM superfluidity is $T_c \sim {\rm mK}$, which intriguingly is comparable to known critical temperatures for cold atom gases, {\it e.g.}, ${}^7$Li atoms have $T_c \simeq 0.2$~mK. We will see that cold atoms provide more than just a useful analogy --- in many ways, our DM component behaves exactly like cold atoms. In cold atom experiments, atoms are trapped using magnetic fields; in our case, it is gravity that attracts DM particles in galaxies.

The superfluid nature of DM dramatically changes its macroscopic behavior in galaxies. Instead of behaving as individual collisionless particles,
the DM is more aptly described as collective excitations, which at low energy are just phonons. In the non-relativistic regime and at lowest order in derivatives,
it is well-known that superfluid phonons are in general described by a scalar field $\theta$ governed by the effective field theory (EFT)~\cite{Son:2002zn}:
\be
{\cal L} = P(X),\; \qquad X =  \dot{\theta}  - m \Phi - \frac{(\vec{\nabla}\theta)^2}{2m}\,,
\label{PXgen}
\ee
where $\Phi$ is the gravitational potential. In particular, the type of superfluid, {\it i.e.}, its equation of state, is uniquely encoded in the choice of $P$. 

Once we take seriously the idea that DM is a superfluid, the only question is --- what kind of superfluid? The second central idea underlying this work is that {\it DM phonons 
are described by the non-relativistic MOND scalar action,}
\be
P(X) \sim \Lambda X\sqrt{|X|}\,.
\label{PMOND}
\ee
where $\Lambda\sim {\rm meV}$ to reproduce the MOND critical acceleration.\footnote{The possible connection between MOND and superfluidity was mentioned briefly by Milgrom in~\cite{Milgrom:2001gz}. We thank A.~Kosowsky for pointing this out to us.} This choice corresponds to a particular superfluid, with $P\sim \rho^3$. To mediate a MONDian force between ordinary matter, phonon must couple to the baryon density:
\be
{\cal L}_{\rm int} \sim \frac{\Lambda}{M_{\rm Pl}} \theta \rho_{\rm b}\,.
\label{Lintintro}
\ee
From a particle physics standpoint, such a coupling is fairly innocuous --- it represents a soft explicit breaking of the global $U(1)$ symmetry.
In the superfluid interpretation, however, where $\theta$ is the phase of a wavefunction, this coupling picks out a preferred phase,
which seems unphysical. One possibility is that~\eqref{Lintintro} follows from baryons coupling to the vortex sector of the superfluid. This would give rise to
a $\cos\theta\rho_{\rm b}$ operator~\cite{Villain,cosinecoupling,Randy}, thereby breaking the continuous shift symmetry down to a discrete subgroup.
When expanded around the state at finite chemical potential $\theta = \mu t$, such operators would give~\eqref{Lintintro} to leading order, albeit with an oscillatory prefactor.

Thus, through~\eqref{PMOND} and~\eqref{Lintro}, phonons play a key role by mediating a long-range force between ordinary matter particles. As a result, a test particle orbiting the galaxy is subject to two forces: the (Newtonian) gravitational force and the phonon-mediated force. Our postulate is that the phonon-mediated force is MONDian, such that the DM superfluid reproduces the empirical success of MOND in galaxies.

The fractional $3/2$ power would be strange if~(\ref{PMOND}) described a fundamental scalar field. 
As a theory of phonons, however, it is not uncommon to see fractional powers in cold atom systems.
For instance, the Unitary Fermi Gas (UFG)~\cite{UFGreview,Giorgini:2008zz}, which has generated much excitement recently in the cold atom community, describes a gas of cold fermionic atoms tuned such that their scattering length diverges~\cite{Feshbach:1962ut,Braaten:2004rn}. The effective action for the UFG superfluid is uniquely fixed by 4d scale invariance at lowest-order in derivatives, ${\cal L}_{\rm UFG}(X) \sim X^{5/2}$, which is also non-analytic~\cite{Son:2005rv}.\footnote{Similarly, in the quasi-static limit ($\dot{\theta} = 0$) our action $\sim X^{3/2}$ becomes invariant under time-dependent spatial Weyl transformations: $h_{ij} \rightarrow \Omega^2(\vec{x},t) h_{ij}$~\cite{Milgrom:1997gx,Milgrom:2008cs}. At lowest order in derivatives it is the unique action with this property. Intriguingly, the $SO(4,1)$ global part of the 3d Weyl group coincides with the de Sitter isometry group, which hints at a deep connection between the MOND phenomenon and dark energy~\cite{Milgrom:2008cs}.}  
 
A hint on the nature of our condensate can be inferred from the (grand canonical) equation of state $P(\mu)$, obtained by working at finite chemical potential $\theta = \mu t$: $P \sim \mu^{3/2}$. Using standard thermodynamics, this implies a polytropic equation of state:
\be
P \sim \rho^3\,.
\ee
We can compare this to the viral expansion $P= k_{\rm B}T \rho + g_2(T)\rho^2 + g_3(T)\rho^3 + \ldots$, where the $\rho$ term describes an ideal gas, the $\rho^2$ term describes 2-body interactions, the $\rho^3$ term 3-body interactions, {\it etc.} The $P\sim \rho^3$ dependence in our case suggests that DM particles have negligible 2-body interactions and interact primarily through 3-body processes. It would be very interesting to find explicit examples of such superfluids in Nature and study in more detail their microphysical interactions.

As is familiar from liquid helium, a superfluid at finite temperature (but below the critical temperature) is best described phenomenologically as a mixture of two fluids~\cite{tisza,london,landau}: $i)$ the superfluid, which by definition has vanishing viscosity and carries no entropy; $ii)$ the ``normal" component, comprised of massive particles, which is viscous and carries entropy.
The fraction of particles in the condensate decreases with increasing temperature. Thus our framework naturally distinguishes between galaxies (where MOND is successful) and galaxy clusters (where MOND is not). Galaxy clusters have a higher velocity dispersion and correspondingly higher DM temperature. For $m\sim {\rm eV}$ we find that galaxies are almost entirely condensed, whereas galaxy clusters are either in a mixed phase or entirely in the normal phase.

Assuming hydrostatic equilibrium with $P \sim \rho^3$, the resulting DM halo density profile is cored, not surprisingly,
and therefore avoids the cusp problem of CDM. Remarkably, for our parameter values ($m\sim {\rm eV}$, $\Lambda \sim {\rm meV}$)
the size of the condensate halo is $\sim 100$~{\rm kpc} for a galaxy of Milky-Way mass. In the inner region of galaxies
where rotation curves are probed, the DM condensate has a negligible effect on baryonic particles, and their motion is dominated by the phonon-mediated MOND force. 
In the outer region probed  by gravitational lensing, the DM condensate gives the dominant contribution to the force on a test particle.

In the vicinity of individual stars the phonon effective theory breaks down and the correct description is in terms of normal DM particles.
This is good news on two counts. First, it is well-known that the MONDian acceleration, while giving a small correction to Newtonian gravity
in the solar system, is typically too large to conform to planetary orbital constraints. This usually requires introducing additional complications to the theory~\cite{Babichev:2011kq}.
In our case, the MONDian behavior is avoided entirely in the solar system, as DM behaves as ordinary particles. The second piece of good news pertains to experimental searches of axion-like particles. By allowing the usual axion-like couplings to Standard Model operators, our DM particles can be detected through the suite of standard axion
experiments, {\it e.g.},~\cite{Essig:2013lka}.

The superfluid interpretation has a number of observational consequences, discussed in detail in Secs.~\ref{bullet}$-$\ref{othercons}, which can potentially
distinguish this scenario from ordinary MOND and $\Lambda$CDM. We mention a few here:

\begin{itemize}

\item As is well-known, a superfluid cannot rotate uniformly; when spun faster than a critical velocity, the superfluid instead develops localized vortices. The
typical angular momentum of galactic haloes is well above the critical velocity, giving rise to an array of DM vortices permeating the galactic disc~\cite{Silverman:2002qx,vortex2}.
Unfortunately these have negligible energy density, so their detection through gravitational lensing may prove challenging. Substructure lensing may soon be possible with the Atacama Large Millimeter Array~\cite{Hezaveh:2012ai}.

\item A key difference with $\Lambda$CDM is the merger rate of galaxies. Applying Landau's criterion for superfluidity, we find two possible outcomes depending on the infall velocity. If the infall velocity is less than the phonon sound speed, then the galactic condensate halos will pass through each other with negligible dissipation. In this case the merger time scale will be much longer than in $\Lambda$CDM and involve multiple encounters, as dynamical friction between the superfluid halos will be negligible. If the infall velocity is greater than the sound speed, the encounter will drive halos out of equilibrium and excite DM particles out of the condensate. In this case dynamical friction will lead to a rapid halo merger, as in $\Lambda$CDM, and after some time the merged halo will thermalize and condense back to the superfluid ground state.

\item The story is even richer for merging galaxy clusters, such as the Bullet Cluster~\cite{Clowe:2003tk,Clowe:2006eq,Bradac:2006er}. Here the outcome not only depends on the infall velocity, but also on the relative fraction of superfluid vs normal components in the clusters. If the infall velocity is sub-sonic, the superfluid components should once again pass through each other with negligible friction, however the normal components should be slowed down due to the significant self-interaction cross section. In general, we therefore expect that lensing maps of bullet-like systems should display two features: $i)$ mass peaks coincident with the cluster galaxies, due to the (non-interacting) superfluid cores; $ii)$ another mass peak coincident with the X-ray luminosity peak, due to the (interacting) normal components.  Remarkably, this picture is consistent with the lensing map of the Abell 520 (MS0451+02) merging system~\cite{Mahdavi:2007yp,Jee:2012sr,Clowe:2012am,Jee:2014hja}. The Bullet Cluster is also consistent with this picture if the sub-cluster (the ``bullet") is predominantly superfluid.

\end{itemize}

The idea of a Bose-Einstein DM condensate (BEC) in galaxies has been studied before~\cite{Silverman:2002qx,vortex2,Boehmer:2007um,Sin:1992bg,Ji:1994xh,Hu:2000ke,Goodman:2000tg,Peebles:2000yy,Arbey:2003sj,Lee:2008ux,Lee:2008jp,Harko:2011xw,Dwornik:2013fra,Guzman:2013rua,Harko:2014vya}.\footnote{In the context of QCD axion, it has been argued that Bose-Einstein condensation can occur in galaxies~\cite{Sikivie:2009qn,Erken:2011dz}, though this has been disputed recently~\cite{Guth:2014hsa}.} There are important differences with the present work. In BEC DM galactic dynamics are caused by the condensate density profile, similar to what happens in CDM, with phonons being irrelevant. In our case, phonons play a key role in generating flat rotation curves and explaining the BTFR. Moreover, the equation of state is different: the BEC DM is governed by two-body interactions and hence has $P \sim \rho^2$, compared to $\sim \rho^3$ in our case. This difference only has a minor effect on the condensate density profiles, but it does imply a different phonon sound speed. In particular, for the Bullet Cluster the sound speed in BEC DM is only $c_s \;\lsim\; 100~{\rm km}/{\rm s}$, {\it i.e.}, more than an order of magnitude smaller than the bullet infall velocity. As a result dissipation is important, which puts BEC DM in tension with observations~\cite{slepiangoodman}.

\section{Dark Matter Condensation}
\label{superfluid}

\begin{figure}[t]
\centering
\includegraphics[width=4in]{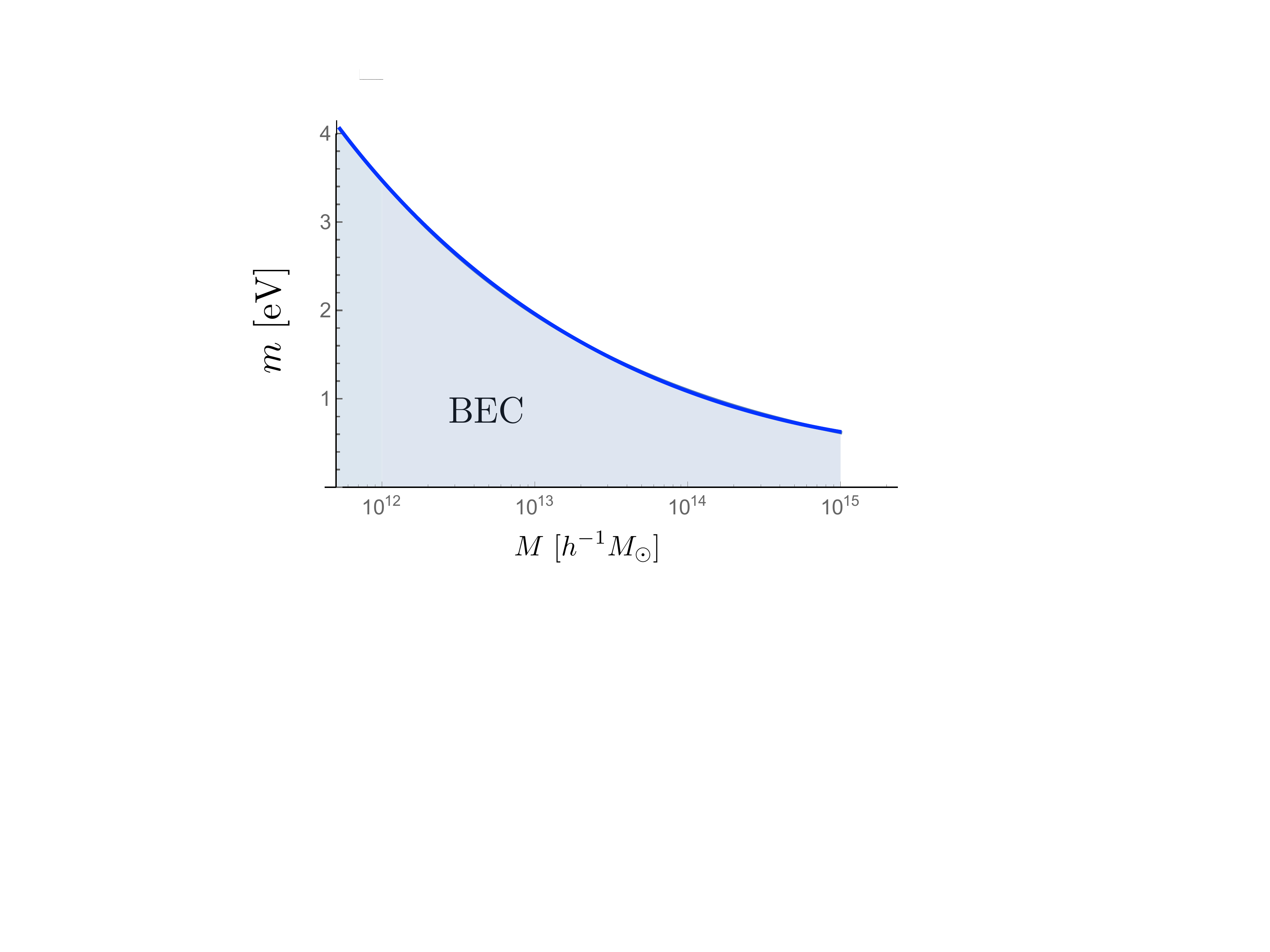}
\caption{\label{massdep} \small Dependence of the BEC region (shaded) on the DM mass $m$ and the halo mass $M$, assuming $z_{\rm vir} = 2$ for concreteness.}
\end{figure}

In order for DM particles to Bose-Einstein condense in galaxies, two conditions must be met. For the purpose
of these initial estimates, we shall treat DM as weakly interacting particles for simplicity, leaving for future
work a refined calculation including interactions. The first condition is that the de Broglie wavelength of DM particles $\lambda_{\rm dB}\sim \frac{1}{mv}$ be larger than the 
mean interparticle separation $\ell\sim \left(\frac{m}{\rho}\right)^{1/3}$. This implies an upper bound on the mass:
\be
m\;\lsim\; \left(\frac{\rho}{v^3}\right)^{1/4}\,.
\label{BECcondbasic}
\ee
We shall apply this bound at virialization, which marks the initial moment when one can meaningfully talk about
an individual halo. From standard collapse theory, virialization occurs when $\frac{\delta\rho}{\rho} \simeq 180$. In terms of
the present DM cosmological density $\rho_{\rm DM}^{(0)}\simeq 3\times 10^{-30}\; {\rm g}/{\rm cm}^3$, the density at virialization is therefore
\be
\rho_{\rm vir} =  (1+z_{\rm vir})^3 \;180 \; \rho_{\rm DM}^{(0)}\simeq (1+z_{\rm vir})^3 \; 5.4\times 10^{-28}\; {\rm g}/{\rm cm}^3\,.
\label{virialrho}
\ee
Meanwhile, the velocity is related to the mass of the object as usual by~\cite{Peacock:1999ye}
\be
v_{\rm vir} = 127\left(\frac{M}{10^{12}h^{-1}M_\odot}\right)^{1/3}\sqrt{1 + z_{\rm vir}}~{\rm km}/{\rm s}\,.
\label{veldispersion}
\ee
Substituting these into~\eqref{BECcondbasic}, we obtain
\be
m ~\lsim~ 2.3 \left(1+z_{\rm vir}\right)^{3/8}\;  \left(\frac{M}{10^{12}h^{-1}M_\odot}\right)^{-1/4}\; {\rm eV}\,.
\ee
Hence light objects form a BEC while heavy objects do not. Figure~\ref{massdep} shows the BEC region as a function of mass assuming $z_{\rm vir} = 2$ for concreteness. 

The second necessary condition for condensation is that DM particles thermalize, with the temperature set by the virial velocity. The interaction rate is given by~\cite{Sikivie:2009qn}
\be
\Gamma\sim {\cal N} v \rho_{\rm vir} \frac{\sigma}{m}\,,
\label{gammaenhanced}
\ee
where 
\be
{\cal N}  \sim \frac{\rho_{\rm vir}}{m} \frac{(2\pi)^3}{\frac{4\pi}{3}(mv)^3} \simeq 10^3 \left(1+z_{\rm vir}\right)^{3/2}\; \;\left(\frac{m}{{\rm eV}}\right)^{-4} \frac{10^{12}h^{-1}M_\odot}{M}
\ee
is the Bose enhancement factor, {\it i.e.}, of order $10^3$ particles for a massive galaxy.\footnote{Strictly speaking,~\eqref{gammaenhanced} is valid provided that $\Gamma \ll mv^2$~\cite{Sikivie:2009qn}, which is easily satisfied in our case.} The interaction rate should be compared to the dynamical time in galaxies, $t_{\rm dyn} \sim \frac{1}{\sqrt{G_{\rm N}\rho_{\rm vir}}}$. 
Indeed, if the time scale for thermalization is shorter than the halo dynamical time, the coherence length for the condensate will be comparable to the size of the halo. This is necessary in order for phonons to act coherently across the galaxy. Putting everything together, the condition $\Gamma t_{\rm dyn} \;\gsim\; 1$ can be expressed as a lower bound on the interaction cross section
\be
\frac{\sigma}{m} \;\gsim\;  \left(1+z_{\rm vir}\right)^{-7/2}\left(\frac{m}{{\rm eV}}\right)^{4} \left(\frac{M}{10^{12}h^{-1}M_\odot}\right)^{2/3}\; 52~\frac{{\rm cm}^2}{{\rm g}}\,.
\ee
Clearly the bound is most stringent for massive galaxies. Taking $M \sim 10^{12}h^{-1}M_\odot$ and assuming $z_{\rm vir} = 2$ for concreteness,
we obtain
\be
\frac{\sigma}{m} \;\gsim\; \left(\frac{m}{{\rm eV}}\right)^{4}\,\frac{{\rm cm}^2}{{\rm g}}\,.
\label{siglow}
\ee
We will see below that a mass of around 0.6~eV gives appropriate size halos, in which case $\frac{\sigma}{m} \;\gsim\; 0.1\;\frac{{\rm cm}^2}{{\rm g}}$. 
The lower end of this bound satisfies current constraints~\cite{MiraldaEscude:2000qt,Gnedin:2000ea,Randall:2007ph} on the cross section of self-interacting dark matter (SIDM)~\cite{Spergel:1999mh}. However, as we will see the phenomenology of superfluid DM is considerably different than SIDM, and each constraint much be carefully revisited. 

\begin{figure}[t]
\centering
\includegraphics[width=5in]{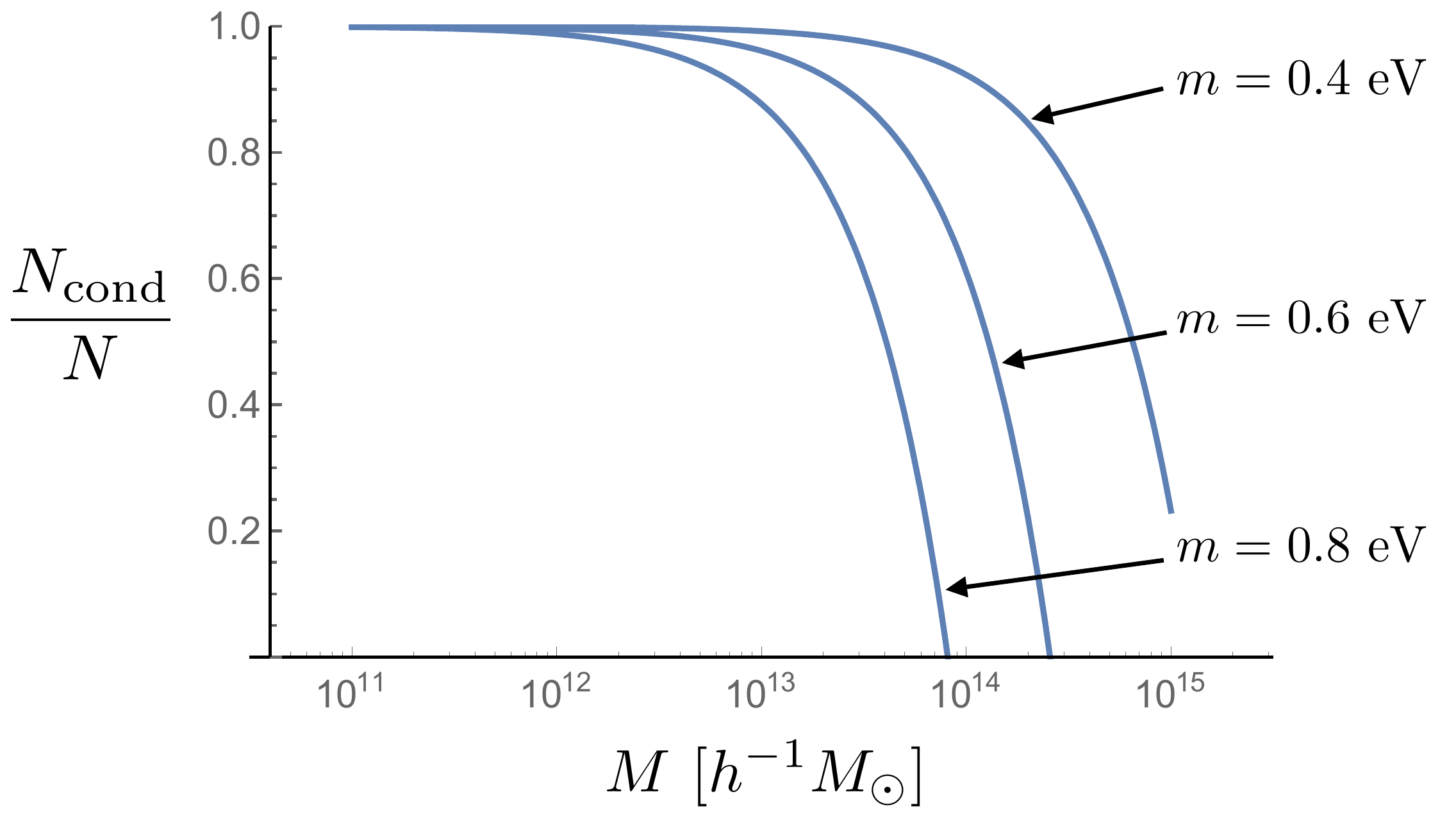}
\caption{\label{Ncond} \small Fraction of DM particles in the condensate as a function of halo mass $M$ for $m=0.4$, $0.6$ and $0.8~{\rm eV}$, assuming $z_{\rm vir} = 0$.}
\end{figure}

The resulting DM temperature is quite cold. The critical temperature can be readily obtained assuming equipartition, $k_{\rm B}T_{\rm c} = \frac{1}{3} mv^2_{\rm c}$,
where $v_{\rm c}$ saturates~\eqref{BECcondbasic}. The result is in the mK range:
\be
T_{\rm c} = 6.5 \;\left(\frac{{\rm eV}}{m}\right)^{5/3} (1+z_{\rm vir})^{2}~{\rm mK}\,.
\ee
The temperature in a given halo, in units of $T_{\rm c}$, is
\be
\frac{T}{T_{\rm c}} \simeq \frac{0.1}{1+z_{\rm vir}} \left(\frac{m}{{\rm eV}}\right)^{8/3}  \left(\frac{M}{10^{12}h^{-1}M_\odot}\right)^{2/3}\,.
\label{ToverTc}
\ee
At finite but sub-critical temperature, the system is phenomenologically described as a mixture of condensate and normal components. Neglecting interactions, the
fraction of condensed particles is~\cite{landaubook}
\bea
\nonumber
\frac{N_{\rm cond}}{N} &=& 1 - \left(\frac{T}{T_{\rm c}}\right)^{3/2}  \\
&\simeq & 1 - \frac{0.03}{(1+z_{\rm vir})^{3/2}}  \left(\frac{m}{{\rm eV}}\right)^{4} \frac{M}{10^{12}h^{-1}M_\odot}\,; \qquad  T \leq T_{\rm c}  \,.
\label{fraccond}
\eea
Figure~\ref{Ncond} plots the condensate fraction as a function of halo mass for $z_{\rm vir} = 0$, for $m = 0.4$, $0.6$ and $0.8~{\rm eV}$. We see that galaxies ($M \;\lsim\; 10^{12}h^{-1}M_\odot$) are almost completely comprised of particles in the condensate, while massive clusters ($10^{14}h^{-1}M_\odot \;\lsim\; M \;\lsim\; 10^{15}h^{-1}M_\odot$) can have a significant fraction, if not all, of their particles in the normal phase. It is worth noting that~\eqref{fraccond} only holds for free particles; one expects the $3/2$ power to change when including interactions. For instance, 
the power is 3 for particles trapped in a harmonic potential. We leave a careful calculation of the condensate fraction including interactions to future work.

A few comments about cosmology. Since our DM particles are in the sub-eV mass range, they are axion-like particles. They must be produced out-of-equilibrium ({\it e.g.} through a phase transition) and remain decoupled from normal matter throughout the history of the universe. For instance, they can be generated through an axion-like vacuum displacement mechanism: in the early universe, the field is displaced from its minimum and starts oscillating once $H\;\lsim\; m$. In this scenario DM particles are generated when $H_i \sim m$. The corresponding photon-baryon temperature is
\be
T_i^{\rm baryons} \sim \sqrt{m M_{\rm Pl}}\,, 
\ee
which for $m\sim {\rm eV}$ is 50~TeV, {\it i.e.} around the weak scale! The velocity is initially relativistic, $v_i \;\lsim\; 1$, and subsequently redshifts as $v\sim 1/a$. 

It is easy to see that, as soon as it generated cosmologically, DM becomes superfluid. Consider the de Broglie wavelength condition~\eqref{BECcondbasic}.
Since $v\sim 1/a$ and $\rho \sim 1/a^3$ cosmologically, both sides of the inequality are time-independent. Hence if~\eqref{BECcondbasic} is satisfied at any time it is satisfied at all times. We can anchor this condition at matter-radiation equality using the observational constraint $\rho_{\rm eq} \simeq 10^{-19}~{\rm g}/{\rm cm}^3 \simeq 0.4~{\rm eV}^4$. Since $v_{\rm eq} \ll 1$, it follows that
\be
m \sim \rho_{\rm eq}^{1/4} \ll  \left(\frac{\rho_{\rm eq}}{v^3_{\rm eq}}\right)^{1/4}\,,
\ee
hence the BEC condition is satisfied at all times. Similarly it is easy to show that thermalization proceeds efficiently, given the lower bound~\eqref{siglow} on $\sigma/m$
and the high occupation number ${\cal N}\gg 1$.  

Naturally DM is much colder on cosmological scales than in collapsed structures. The temperature ratio $T/T_{\rm c} = (v/v_{\rm c})^2$ is constant cosmologically, where $v_c  \equiv (\rho/m^4)^{1/3}$ saturates~\eqref{BECcondbasic}. Once again it is convenient to evaluate this at matter-radiation equality:
\be
\left(\frac{T}{T_{\rm c}}\right)_{\rm cosmo} \simeq v_{\rm eq}^2 \left(\frac{m}{{\rm eV}}\right)^{8/3}\,.
\ee
Assuming $v_i \sim 1$ when $T_i^{\rm baryons} \sim \sqrt{m M_{\rm Pl}}$, we have $v_{\rm eq}  = v_i \frac{a_i}{a_{\rm eq}} \simeq \frac{{\rm eV}}{\sqrt{mM_{\rm Pl}}}$,
and therefore
\be
\left(\frac{T}{T_{\rm c}}\right)_{\rm cosmo} \simeq 10^{-28} \left(\frac{m}{{\rm eV}}\right)^{5/3}\,,
\ee
which is very cold indeed. In contrast we see from~\eqref{ToverTc} that $T/T_{\rm c}$ ranges from $10^{-6}$ in dwarf galaxies ($M \sim 10^{6}\; M_\odot$) to $10^{-2}$ in massive galaxies ($M\sim 10^{12}\; M_\odot$). In other words, cosmologically the DM superfluid can be described to an excellent approximation as a $T=0$ superfluid. In collapsed structures, finite-temperature effects can 
be significant. As we will see, finite-temperature effects will be important in stabilizing the MOND phenomenon in galaxies.

\section{Superfluid Phase}
\label{superphase}

Once DM condenses and forms a superfluid, the relevant low-energy degrees of freedom are collective excitations in the form of phonons.
Superfluid phonons are the Goldstone bosons for a spontaneously broken global $U(1)$ symmetry. In the non-relativistic regime they are in
general described by a scalar field $\theta$ with effective action~\cite{Son:2005rv} 
\be
{\cal L} = P(X) \,;\qquad X = \dot{\theta} - m \Phi - \frac{(\vec{\nabla}\theta)^2}{2m}\,,
\label{Lphongen}
\ee
where $\Phi$ is the external gravitational potential, {\it e.g.},  $\Phi (r)  = - \frac{G_{\rm N}M(r)}{r}$ for a
spherical-symmetric static source. This effective Lagrangian is {\it exact} at lowest order in derivatives, with corrections suppressed by
additional derivatives per field. To describe phonons at constant chemical potential~$\mu$, we expand 
\be
\theta = \mu t + \phi~~\Longrightarrow~~ X = \mu - m \Phi + \dot{\phi} - \frac{(\vec{\nabla}\phi)^2}{2m}\,.
\label{finitemu}
\ee

In the case of interest, our conjecture is that the DM superfluid phonons are governed by the MOND action~\eqref{PMOND},
\be
P(X) =  \frac{2\Lambda(2m)^{3/2}}{3}X\sqrt{|X|} \,.
\label{Lphon}
\ee
The square-root form is necessary to ensure that the action makes sense for time-like field profiles and that the Hamiltonian
is bounded from below~\cite{Bruneton:2007si}. Note that the effective action~\eqref{Lphon} is only well-defined {\it away from} $X = 0$, for both
time-like and space-like profiles. In Sec.~\ref{phi6} we will give a more fundamental derivation of the phonon action starting from a
complex scalar field with $|\partial \Psi|^6$ interactions. As we will see, in that example a condensate only forms for
$2m|X| > \frac{\Lambda_{\rm c}^4}{\Lambda^2}$, for some cutoff scale $\Lambda_{\rm c}$. 

To mediate a MOND force, phonons must couple to the baryon mass density $\rho_{\rm b}$:
\be
{\cal L}_{\rm int} = -\alpha \frac{\Lambda}{M_{\rm Pl}} \theta \rho_{\rm b} \,,
\label{Lint}
\ee
where $\alpha$ is a dimensionless parameter. (The relativistic extension is more complicated and will be discussed in Sec.~\ref{lensing}.) 
This operator explicitly breaks the shift symmetry only at the $1/M_{\rm Pl}$ level and is therefore technically natural.
From the superfluid perspective,~\eqref{Lint} can arise if baryonic matter couple to the vortex sector of the superfluid, giving rise to
operators $\sim \cos\theta \rho_{\rm b}$ that preserve a discrete subgroup of the continuous shift symmetry~\cite{Villain,cosinecoupling,Randy}.
Expanding around a state at finite chemical potential, $\phi = \theta - \mu t$, this operator would yield a coupling of the form~\eqref{Lint}
with an oscillatory prefactor.  For the purpose of the present work we shall treat~\eqref{Lint} as an empirical term in our action necessary to obtain the MOND phenomenon. 

To summarize, our phonon theory depends on three parameters: the particle mass $m$, the scale $\Lambda$ and the coupling constant $\alpha$.
The latter two parameters can depend on temperature, and thus on velocity, most naturally through the ratio $T/T_{\rm c}$.
In particular they can assume different values on cosmological scales (where $T/T_{\rm c}\sim 10^{-28}$) than in galaxies
(where $T/T_{\rm c}\sim 10^{-6}-10^{-2}$). Specifically we will see in Sec.~\ref{cosmo} that $\alpha$ must be $\sim 10^{-4}$ smaller cosmologically,
while $\Lambda$ must be $\sim 10^4$ larger, in order to obtain an acceptable cosmology. The temperature dependence is therefore quite mild
and can be ignored over the velocity range spanned by galaxies. Until Sec.~\ref{cosmo} it will be implicitly understood that $\alpha$ and $\Lambda$ assume
their galactic values, ignoring any temperature dependence. For galaxy phenomenology, we will find in Sec.~\ref{MOND} that these two parameters must be
related in order to reproduce the MOND critical acceleration:
\be
\alpha^{3/2} \Lambda = \sqrt{a_0 M_{\rm Pl}}\simeq 0.8~{\rm meV} ~~\Longrightarrow ~~\alpha \simeq 0.86 \left( \frac{\Lambda}{{\rm meV}}\right)^{-2/3}\,.
\label{LambdaMOND}
\ee
Hence $\alpha\sim {\cal O}(1)$ for $\Lambda\sim {\rm meV}$.

\subsection{Condensate and phonon properties}

The form of the phonon action~\eqref{Lphon} uniquely fixes the properties of the condensate through standard thermodynamics arguments.
We work at finite chemical potential, $\theta = \mu t$, setting phonon excitations and gravitational potential to zero. The pressure of the condensate is given as usual by the Lagrangian density,
\be
P(\mu) = \frac{2\Lambda}{3} (2m\mu)^{3/2}\,.
\ee
This is the grand canonical equation of state $P = P(\mu)$ for the condensate. Differentiating with respect to $\mu$ yields the number density of condensed particles:
\be
n = \frac{\partial P}{\partial\mu} = \Lambda (2m)^{3/2} \mu^{1/2}\,.
\label{nmu}
\ee
Combining these expressions and using the non-relativistic relation $\rho= m n$, we find
\be
P = \frac{\rho^3}{12\Lambda^2m^6}\,.
\label{eos}
\ee
This is a polytropic equation of state $P \sim \rho^{1 + 1/n}$ with index $n = 1/2$. In comparison, the standard DM BEC discussed 
in the literature is described by $P\sim \rho^2$, corresponding to $n=1$. We will see below that the halo profiles are nonetheless quite similar. 

Let us now consider phonon excitations on top of this condensate. Expanding~\eqref{Lphon} to quadratic order in phonon perturbations $\phi = \theta - \mu t$, once again neglecting the gravitational potential,
we obtain
\be
{\cal L}_{\rm quad} =  \frac{\Lambda (2m)^{3/2}}{4\mu^{1/2}} \left( \dot{\phi}^2 - \frac{2\mu}{m} (\vec{\nabla}\phi)^2\right)\,.
\ee
The sound speed is
\be
c_s = \sqrt{\frac{2\mu}{m}}\,.
\label{cs}
\ee
Expanding to higher order, we can identify the strong coupling scale of the theory. A typical interaction term is schematically of the form
\be
{\cal L}_{\rm higher-order} \supset \Lambda m^{3/2} \mu^{3/2-n} \partial^n \phi^n \sim  \left(\Lambda m^{3/2}\mu^{3/2}\right)^{1 - \frac{n}{2}}  \partial^n \phi^n_c\,,
\ee
where $\partial$ stands for either $\partial_t$ or $c_s \vec{\nabla}$, and the canonical variable is $\phi_c \sim \Lambda^{1/2}m^{3/4} \mu^{-1/4}\phi$.
The strong coupling scale, identified as the scale suppressing higher-dimensional operators, is
\be
\Lambda_{\rm s} = \left(\Lambda m^{3/2}\mu^{3/2}\right)^{1/4}\,.
\label{Ls}
\ee

\subsection{Halo profile}
\label{haloprof}

Given the equation of state~\eqref{eos}, we can compute the DM density profile of the condensate halo
assuming hydrostatic static equilibrium. Focusing on a static, spherically-symmetric halo, the pressure
and acceleration are related by
\bea
\nonumber
\frac{1}{\rho(r)}\frac{{\rm d}P(r)}{{\rm d}r} &=& - \frac{{\rm d}\Phi(r)}{{\rm d}r} \\
&=&  - \frac{4\pi G_{\rm N}}{r^2} \int_0^r {\rm d}r' r'^2\rho(r')\,.
\label{hydro}
\eea
Equivalently, since by definition $\rho = m n = m \frac{{\rm d}P}{{\rm d}X}$ this equation can be written as
\be
\frac{{\rm d}X(r)}{{\rm d}r} = - m \frac{{\rm d}\Phi(r)}{{\rm d}r}\,,
\label{dXdr}
\ee
which automatically follows from the expression~\eqref{finitemu} for $X$ when phonon excitations are set to zero.

\begin{figure}[t]
\centering
\includegraphics[width=4.5in]{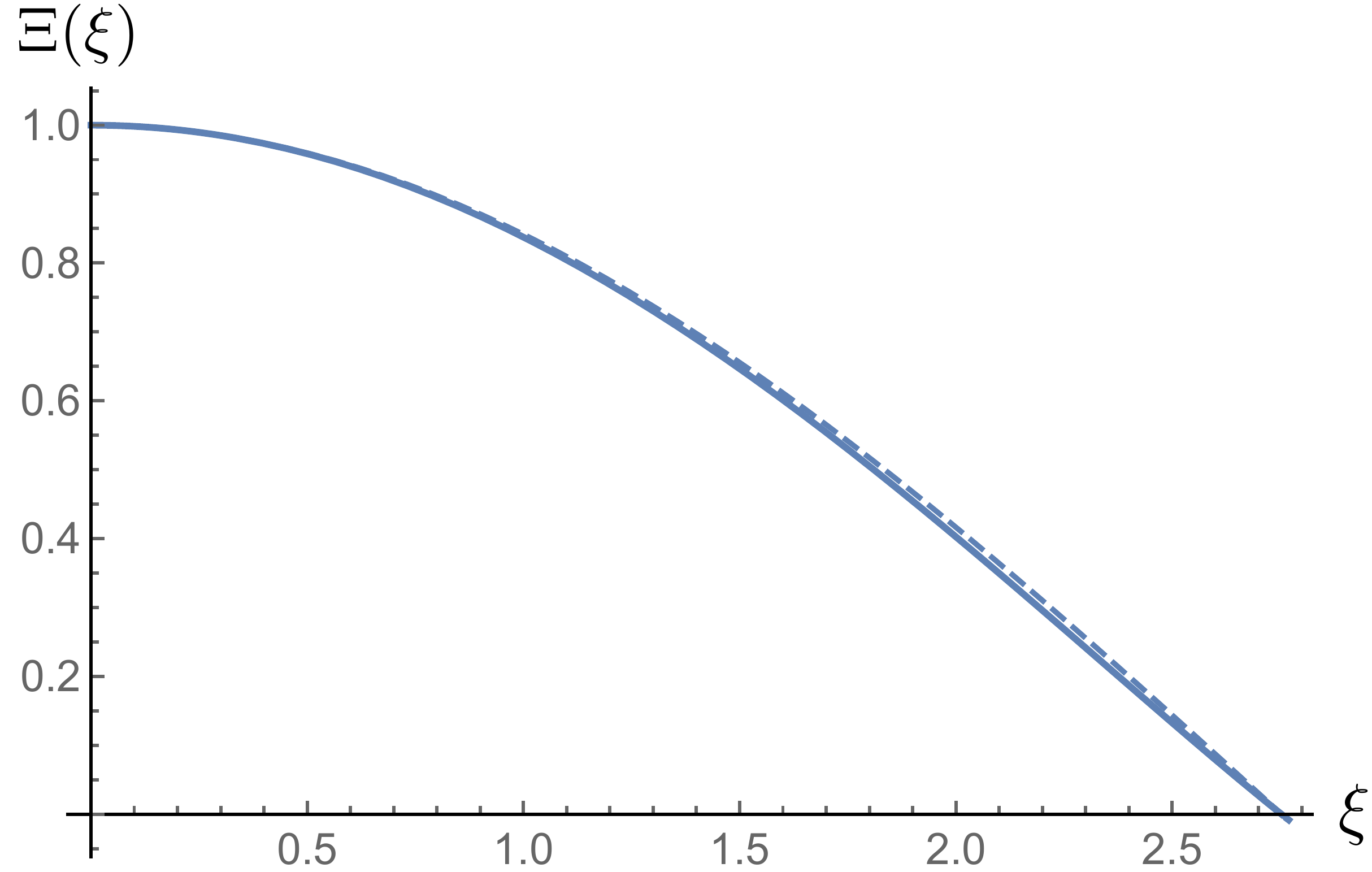}
\caption{\label{laneemden} \small Numerical solution to the $n=1/2$ Lane-Emden equation with boundary condition $\Xi(0) = 1$ and $\Xi'(0) = 0$. The solution vanishes at $\xi_1 \simeq 2.75$. The dashed line is a simple approximate analytical form, $\Xi(\xi) = \cos\left(\frac{\pi}{2}\frac{\xi}{\xi_1}\right)$.}
\end{figure}

It is convenient to rewrite this equation in terms of dimensionless variables $\Xi$ and $\xi$, defined by
\bea
\nonumber
\rho (r) &=&\rho_0\Xi^{1/2} \,;\\
r &=& \sqrt{\frac{\rho_0}{32\pi G_{\rm N} \Lambda^2 m^6}}~\xi\,,
\eea
where $\rho_0 \equiv \rho(0)$ is the central density. Differentiating~\eqref{hydro} with respect to $r$, and expressing the result in the new variables, it is straightforward to obtain
the $n = 1/2$ Lane-Emden equation:\footnote{The Lane-Emden equation for general $n$ is
\be
\frac{1}{\xi^2}\frac{{\rm d}}{{\rm d}\xi} \left(\xi^2 \frac{{\rm d}\Xi}{{\rm d}\xi}\right) = - \Xi^n\,.
\ee
Analytical solutions exist for $n=0$, 1 and 5~\cite{chandrabook}. Other values of $n$ require numerical integration.}
\be
\frac{1}{\xi^2}\frac{{\rm d}}{{\rm d}\xi} \left(\xi^2 \frac{{\rm d}\Xi}{{\rm d}\xi}\right) = - \Xi^{1/2}\,.
\label{LaneEmdenEqn}
\ee
The boundary conditions are $\Xi(0) = 1$ and $\Xi'(0) = 0$. The numerical solution, shown in Fig.~\ref{laneemden},
vanishes at 
\be
\xi_1 \simeq 2.75\,, 
\ee
which defines the size of the condensate:
\be
R = \sqrt{\frac{\rho_0}{32\pi G_{\rm N} \Lambda^2 m^6}}\; \xi_1 \,.
\label{Rhalo}
\ee

A simple analytical form that provides a good fit is $\Xi(\xi) = \cos\left(\frac{\pi}{2}\frac{\xi}{\xi_1}\right)$,
shown as the dashed curve in the Figure. The density profile is thus well approximated by
\be
\rho(r)\simeq \rho_0 \cos^{1/2}\left(\frac{\pi}{2} \frac{r}{R}\right) \,;\qquad r \leq R\,.
\label{rhoapprox}
\ee
The central density is related to the mass of the halo condensate as follows~\cite{chandrabook}
\be
\rho_0 =  \frac{M}{4\pi R^3} \frac{\xi_1}{|\Xi'(\xi_1)|} \,.
\ee
From the numerics we find $\Xi'(\xi_1) \simeq -0.5$. Substituting~\eqref{Rhalo}, we can solve for the central density
\be
\rho_0 \simeq \left(\frac{M_{\rm DM}}{10^{12}M_\odot}\right)^{2/5} \left(\frac{m}{{\rm eV}}\right)^{18/5} \left(\frac{\Lambda}{{\rm meV}}\right)^{6/5}  \; 10^{-24}~{\rm g}/{\rm cm}^3\,.
\label{rhocentral}
\ee
Meanwhile the halo radius is 
\be
R \simeq \left(\frac{M_{\rm DM}}{10^{12}M_\odot}\right)^{1/5} \left(\frac{m}{{\rm eV}}\right)^{-6/5}\left(\frac{\Lambda}{{\rm meV}}\right)^{-2/5} \; 45~{\rm kpc}\,.
\label{halorad}
\ee

Remarkably, for $m\sim {\rm eV}$ and $\Lambda\sim {\rm meV}$ we obtain DM halos of realistic size.  In the standard CDM picture a halo of mass $M_{\rm DM} = 10^{12}\,M_\odot$ has a virial radius of $\sim 200$~kpc. In our framework, the condensate radius can in principle be considerably smaller or larger depending on parameter values. For concreteness, in the remainder of the analysis we will choose as fiducial values
\be
m = 0.6~{\rm eV}\,;\qquad \Lambda = 0.2~{\rm meV}\,.
\label{fidparam}
\ee
(From~\eqref{LambdaMOND} this corresponds to $\alpha\simeq 5/2$.) This implies a condensate radius of $\sim 158$~kpc for a halo of mass $M_{\rm DM} = 10^{12}\,M_\odot$. 

Through the relation $\rho = m n = m \frac{{\rm d}P}{{\rm d}X}$, the above density profile fixes $X(r)$:
\bea
\nonumber
X(r) &=& \frac{\rho^2}{8\Lambda^2m^5} \\
&\simeq & 2\times 10^{-6}~{\rm eV} \left(\frac{M_{\rm DM}}{10^{12}M_\odot}\right)^{4/5} \left(\frac{m}{{\rm eV}}\right)^{11/5}\left(\frac{\Lambda}{{\rm meV}}\right)^{2/5}  \cos\left(\frac{\pi}{2} \frac{r}{R}\right)\,.
\label{mur}
\eea
In particular, the central density determines the chemical potential:
\be
\mu =  \frac{\rho^2_0}{8\Lambda^2m^5} \,,
\label{mualone}
\ee
which in turns determines the strong coupling scale~\eqref{Ls}:
\bea
\nonumber
\Lambda_{\rm s} &=& \frac{\rho^{3/4}_0}{8^{3/8}\Lambda^{1/2} m^{3/2}} \\
&\simeq & {\rm meV} \left(\frac{M_{\rm DM}}{10^{12}M_\odot}\right)^{3/10} \left(\frac{m}{{\rm eV}}\right)^{6/5}\left(\frac{\Lambda}{{\rm meV}}\right)^{2/5} \,,
\label{Lsexplicit}
\eea
Thus the strong coupling scale, like $\Lambda$, is of order meV. Finally, the gravitational potential $\Phi(r)  = m^{-1} \left(X(r) -\mu\right)$ follows trivially from these relations.

A few comments are in order. First we have neglected the effect of halo rotation in this calculation. Slowly-rotating BEC with polytopic equation of state can incorporated into a modified Lane-Emden equation~\cite{chandrapaper}. However we will see in Sec.~\ref{vortices} that rotating halos are typically unstable to the formation of quantum vortices, which carry the angular momentum. Second, $R$ represents the size of the superfluid ``core", not of the entire halo. In reality we expect this core to be surrounded by DM particles in the normal phase, most likely described by a Navarro-Frenk-White (NFW) profile~\cite{Navarro:1996gj}. A careful analysis would require numerical simulations, which is beyond the scope of this paper. Third, the superfluid scenario offers a simple, if not mundane, resolution to the cusp-core and ``too big to fail" problems~\cite{BoylanKolchin:2011de,toobigtofail2}. The density profile is cored and hence has a much lower central density than in collisionless CDM simulations, in better agreement with the inferred densities of MW dwarf satellites. 

\section{Including Baryons: Phonon-Mediated Force}
\label{MOND}

In this Section we derive the phonon profile in galaxies, modeling the baryons as a static, spherically-symmetric localized source for simplicity.
We first focus on the zero-temperature analysis, where the Lagrangian is given by the sum of~\eqref{Lphon} and~\eqref{Lint}. In this case we
find two branches of solutions, depending on the sign of $X$. The branch with $X>0$ has stable perturbations but does not admit a
MONDian regime. The branch with $X< 0$ does admit a MONDian regime, where the phonon-mediated force approximates the MOND
force law over the scales probed by galactic rotation curve observations, as desired. However, perturbations on this branch are unstable.
Stability on the MOND branch can be restored by finite-temperature effects, as we will show in Sec.~\ref{2fluid}.

\subsection{Zero-temperature analysis}

Recall our zero-temperature phonon Lagrangian:
\be
{\cal L} =  \frac{2\Lambda(2m)^{3/2}}{3}X\sqrt{|X|} - \alpha \frac{\Lambda}{M_{\rm Pl}} \theta \rho_{\rm b}\,.
\label{zeroTLag}
\ee
In the static spherically-symmetric approximation, $\theta = \mu t + \phi(r)$, the equation of motion reduces~to
\be
\vec{\nabla} \cdot \left( \sqrt{2m|X|}~\vec{ \nabla}\phi\right) = \frac{\alpha\rho_{\rm b}(r)}{2M_{\rm Pl}}\,,
\ee
where $X(r) = \mu -m\Phi(r) - \frac{\phi'^2(r)}{2m}$. This can be readily integrated:
\be
\sqrt{2m|X|}~\phi' = \frac{\alpha M_{\rm b}(r)}{8\pi M_{\rm Pl} r^2}\equiv \kappa(r)\,.
\label{kappadef}
\ee
The profile depends on the sign of $X$:

\begin{itemize}

\item {\bf $X > 0$ branch}: In this case the solution is
\be
\phi'(r) = \sqrt{m}\left( \hat{\mu}   - \sqrt{\hat{\mu}^2 -  \frac{\kappa^2}{m^2}}\right)^{1/2}\,; \qquad \hat{\mu} \equiv \mu - m\Phi\,,
\label{phigensoln1}
\ee
where we have chosen the minus sign such that $\phi'\rightarrow 0$ when $M_{\rm b} \rightarrow 0$. Equivalently, the solution for $X(r)$ is
\be
X(r) = \frac{1}{2} \left( \hat{\mu}+ \sqrt{\hat{\mu}^2 - \frac{\kappa^2}{m^2}}\right)\,.
\label{X>0soln}
\ee
As a check note that $X\rightarrow \hat{\mu}$ for $M_{\rm b} \rightarrow 0$, which is consistent with our equation~\eqref{dXdr} for the density profile
in the absence of baryons. More generally, we can solve~\eqref{X>0soln} for the gravitational potential: $\hat{\mu} = \mu - m\Phi = X + \frac{\kappa^2}{4m^2 X}$.
Substituting into Poisson's equation, we obtain
\be
\nabla^2\left(X + \frac{\kappa^2}{4m^2 X}\right) = - \frac{m^4\Lambda}{M_{\rm Pl}^2} \left(\frac{X}{m}\right)^{1/2} + \frac{\rho_{\rm b}}{2M_{\rm Pl}}\,,
\label{genPoissonX>0}
\ee
where we have used $\rho = m\frac{{\rm d}P}{{\rm d}X}$ for the condensate matter density. In the absence of baryons, this reduces to the Lane-Emden equation~\eqref{LaneEmdenEqn}. In the presence of baryons,
it is easy to show that the solution is qualitatively similar, with the only notable difference being that the halo radius shrinks with increasing baryonic mass, as expected from the extra gravitational attraction due to baryons. 

\item {\bf $X < 0$ branch}: On this branch the solution is
\be
\phi'(r) = \sqrt{m}\left(\hat{\mu}  + \sqrt{\hat{\mu}^2 +  \frac{\kappa^2}{m^2}}\right)^{1/2}\,,
\label{phigensoln}
\ee
where we have dismissed a solution corresponding to imaginary $\phi'$.  Equivalently, the solution for $X(r)$ is
\be
X(r) = \frac{1}{2} \left( \hat{\mu} - \sqrt{\hat{\mu}^2 + \frac{\kappa^2}{m^2}}\right)\,.
\label{X<0soln}
\ee
Unlike the $X>0$ solution, this branch admits a MONDian regime where $\kappa \gg \hat{\mu}$, such that
\be
\phi'(r) \simeq  \sqrt{\kappa(r)} = \sqrt{\frac{\alpha M_{\rm b}(r)}{8\pi M_{\rm Pl} r^2}} \,.
\label{phiMOND}
\ee
In this limit the scalar acceleration on an ordinary matter particle is
\be
a_\phi(r) = \alpha \frac{\Lambda}{M_{\rm Pl}} \phi' \simeq \sqrt{\frac{\alpha^3\Lambda^2}{M_{\rm Pl}}\frac{G_{\rm N} M_{\rm b}(r)}{r^2} }\,.
\label{aphi}
\ee
To reproduce the MONDian result $a_{\rm MOND}  = \sqrt{a_0 \frac{G_{\rm N} M_{\rm b}(r)}{r^2}}$, we are therefore led to identify
\be
\alpha^{3/2} \Lambda = \sqrt{a_0 M_{\rm Pl}}\simeq 0.8~{\rm meV} ~~\Longrightarrow ~~\alpha \simeq 0.86 \left( \frac{\Lambda}{{\rm meV}}\right)^{-2/3}\,,
\label{alphasoln}
\ee
which fixes $\alpha$ in terms of $\Lambda$ through the critical acceleration, as claimed earlier. That $\Lambda$ is of order the dark energy scale is a
direct consequence of the coincidence $a_0 \sim H_0$.

Repeating the steps that led to~\eqref{genPoissonX>0}, in this case we find
\be
\nabla^2\left(X - \frac{\kappa^2}{4m^2 X}\right) = - \frac{m^4\Lambda}{M_{\rm Pl}^2} \left(\frac{-X}{m}\right)^{1/2} + \frac{\rho_{\rm b}}{2M_{\rm Pl}}\,.
\label{genPoissonX<0}
\ee
This equation generically leads to unphysical halos, with growing DM density as a function of $r$. The origin of this instability can be seen at the level of perturbations.
Expanding~\eqref{zeroTLag} to quadratic order in phonon perturbations $\varphi = \phi - \bar{\phi}(r)$, we obtain
\be
{\cal L}_{\rm quad} = {\rm sign}(\bar{X}) \frac{\Lambda(2m)^{3/2}}{4\sqrt{|\bar{X}|}} \left( \dot{\varphi}^2 - 2\frac{\bar{\phi}'}{m} \varphi'\dot{\varphi} - 2\frac{\varphi'^2}{m}\left(\bar{X} - \frac{\bar{\phi}'^2}{2m}\right) - \frac{2\bar{X}}{mr^2} (\partial_\Omega\varphi)^2 \right)\,.
\label{LMONDquad}
\ee
The kinetic term $\dot{\varphi}^2$ has the wrong sign for $\bar{X} < 0$. 

\end{itemize}

To summarize, the $X> 0$ solution, given by~\eqref{phigensoln1}, is continuously connected to the homogeneous condensate in the absence of baryons ($M_{\rm b}\rightarrow 0$) and has stable perturbations. However, 
this branch does not admit a MONDian regime. The $X< 0$ solution, on the other hand, does admit an approximate MOND regime, but this branch has the peculiarity that $\phi'$ remains non-zero
even in the $M_{\rm b}\rightarrow 0$ limit. Moreover, perturbations about this solution have wrong-sign kinetic term, indicating an instability. Below we will show that this instability can be cured by finite-temperature effects.

\subsection{Finite-temperature effects}
\label{2fluid}

The DM condensate in actual galactic halos has non-zero temperature, hence we expect that the zero-temperature Lagrangian~\eqref{zeroTLag}
receives finite-temperature corrections in galaxies. At finite sub-critical temperature, the system is described phenomenologically by Landau's two-fluid model: an admixture of a superfluid component, which has zero viscosity, and a normal component, which is viscous and carries entropy. The two components interact with each other. Their relative fraction is a function of temperature,
and hence the mass of the collapsed object, as sketched in Fig.~\ref{Ncond}. 

At lowest order in derivatives, the effective field theory at finite temperature and finite chemical potential is~\cite{Nicolis:2011cs}
\be
{\cal L}_{T \neq 0} = F(X,B,Y)\,.
\ee
It is a function of three scalar quantities. The scalar $X$, already defined in~\eqref{Lphongen}, describes the phonon excitations. The remaining scalars
are defined in terms of the three Lagrangian coordinates $\psi^I(\vec{x},t)$, $I = 1,2,3$ of the normal fluid:\footnote{In~\cite{Nicolis:2011cs}, $Y$ is defined in terms of the relativistic phonon field $\Theta$ as $Y = u^\mu\partial_\mu \Theta$. To translate to the non-relativistic description, we have substituted $\Theta = mt + \theta$ and subtracted the mass term.}
\bea
\nonumber
B &\equiv& \sqrt{{\rm det}\;\partial_\mu\psi^I\partial^\mu\psi^J} \;; \\
Y &\equiv&  u^\mu\left(\partial_\mu\theta + m\delta_\mu^{\;0}\right) -m \simeq  \mu - m\Phi +  \dot{\phi} + \vec{v}\cdot \vec{\nabla}\phi \,,
\label{BY}
\eea
where $u^\mu = \frac{1}{6B} \epsilon^{\mu\alpha\beta\gamma}\epsilon_{IJK}\partial_\alpha\psi^I\partial_\beta\psi^J \partial_\gamma\psi^K$
is the unit 4-velocity vector, and in the last step for $Y$ we have taken the non-relativistic limit $u^\mu \simeq (1-\Phi, \vec{v})$. By construction, these scalars respect the internal symmetries: $i)$ $\psi^I \rightarrow \psi^I + c^I$ (translations); $ii)$ $\psi^I  \rightarrow R^I_{\;J} \psi^J$ (rotations); $iii)$ $\psi^I \rightarrow \xi^I(\psi)$,
with $\det \frac{\partial\xi^I}{\partial\psi^J} = 1$ (volume-preserving reparametrizations).

Our goal is to seek a finite-temperature theory that will generate a MONDian phonon profile~\eqref{phiMOND} over the scales probed by galactic rotation curve observations, while having stable perturbations and
reasonable DM density profile. There is much freedom in specifying finite-temperature operators that will do the trick. The simplest possibility is to supplement~\eqref{zeroTLag} with the
two-derivative operator
\be
\Delta {\cal L} = M^2Y^2 = M^2\left(\mu - m\Phi +  \dot{\phi} \right)^2\,,
\ee
where in the last step we have specialized to the normal fluid rest frame, $\vec{v} = 0$. This leaves the static profile~\eqref{phigensoln} unchanged, however it does modify the quadratic Lagrangian~\eqref{LMONDquad} by an amount $\Delta {\cal L}_{\rm  quad} = M^2 \dot{\varphi}^2$. This will flip the sign of the kinetic term, and therefore cure the ghost, if
\be
M \;\gsim\; \frac{\Lambda m^{3/2}}{\sqrt{|\bar{X}|}} \sim 0.5~\left(\frac{10^{11}\,M_\odot}{M_{\rm b}}\right)^{1/4}  \left(\frac{\Lambda}{{\rm meV}}\right)^{1/2}  \left(\frac{r}{10~{\rm kpc}}\right)^{1/2} \;m \,,
\label{Mbound}
\ee
which, remarkably, is of order eV\;! Hence, for quite natural values of $M$, this two-derivative operator can restore stability. 
Furthermore, this operator gives a contribution $\Delta P = M^2 \mu^2$ to the condensate pressure, which obliterates the unwanted
growth in the DM density profile mentioned below~\eqref{genPoissonX<0}. Instead, the pressure is positive far from the baryons, resulting
in localized, finite-mass halos.

As another example, consider the finite-temperature Lagrangian:
\be
P(X,T) =   \frac{2\Lambda(2m)^{3/2}}{3}X\sqrt{\left\vert X- \beta Y \right\vert} = \frac{2\Lambda(2m)^{3/2}}{3}X\sqrt{\left\vert X- \beta \left(\mu - m\Phi +  \dot{\phi} \right)\right\vert}\,,
\label{LphonT}
\ee
where we have once again focused on the normal fluid rest frame. The dimensionless $\beta$ parameter implicitly depends (mildly) on $T/T_{\rm c}$, though
we will treat it henceforth as constant. This is of course a more {\it ad hoc} form of finite-temperature effects, but it has the advantage of facilitating the analysis.
As we will see, in order to reproduce the MOND phenomenon with stable perturbations we will need
\be
\beta \geq \frac{3}{2}\,, 
\ee
in which case the quantity within absolute values is negative definite.

First consider the DM density profile in the absence of baryons. Setting phonons and gravitational potential to zero, the pressure of the condensate is now given by
\be
P(\mu,T) = \frac{2\sqrt{\beta -1}\Lambda}{3} (2m\mu)^{3/2}\,.
\label{Pbeta}
\ee
Thus the density profile is identical to the zero-temperature profile described in Sec.~\ref{haloprof}, modulo the replacement $\Lambda\rightarrow \sqrt{\beta-1}\Lambda$.
For instance, instead of~\eqref{halorad} the halo radius is now given by
\be
R(T) \simeq \left(\frac{M_{\rm DM}}{10^{12}M_\odot}\right)^{1/5} \left(\frac{m}{{\rm eV}}\right)^{-6/5}\left(\frac{\Lambda}{{\rm meV}}\right)^{-2/5}(\beta-1)^{-1/5} \; 45~{\rm kpc}\,.
\label{haloradT}
\ee

Including baryons, the static, spherically-symmetric scalar equation becomes
\be
\vec{\nabla} \cdot \left( \frac{\phi'^2 + 2m\left(\frac{2\beta}{3} -1\right)\hat{\mu}}{\sqrt{\phi'^2 + 2m(\beta -1)\hat{\mu}}}~\vec{ \nabla}\phi\right) =\frac{\alpha\rho_{\rm b}(r)}{2M_{\rm Pl}}\,, 
\ee
where $\hat{\mu} = \mu- m\Phi$ was introduced in~\eqref{phigensoln1}. This integrates to
\be
\frac{\phi'^2 + 2m\left(\frac{2\beta}{3} -1\right)\hat{\mu}}{\sqrt{\phi'^2 + 2m(\beta -1)\hat{\mu}}}~\phi' =\kappa(r) \,.
\label{neweom}
\ee
This implies a cubic equation for $\phi'^2$, whose real root does not have a particularly illuminating analytic form. For concreteness we shall assume that $\beta$ is strictly greater than $3/2$. The solution then has the following behavior: sufficiently close to the baryon source, such that $\phi'^2 \gg m\hat{\mu}$, the solution approximates the MOND profile~\eqref{phiMOND}, $\phi'\simeq \sqrt{\kappa}$, and therefore scales as $1/r$. Far from the baryons, such that $\phi'^2 \ll m\hat{\mu}$, the solution tends to $\phi' \simeq \sqrt{\frac{3}{2m\hat{\mu}}}\sqrt{\frac{\beta-1}{2\beta-3}} \kappa$, which approximately scales as $1/r^2$ since $\hat{\mu}$ is approximately constant. To summarize, assuming $\beta > 3/2$ the phonon profile is given by
\be
\phi' \simeq \left\{\begin{array}{cl}
\sqrt{\kappa} \sim \frac{1}{r} \hspace{20pt}&\text{if}\hspace{10pt} r \ll r_\star\,, \\ \\
 \sqrt{\frac{3}{2m\hat{\mu}}}\sqrt{\frac{\beta-1}{2\beta-3}} \kappa\sim \frac{1}{r^2} \hspace{20pt}&\text{if}\hspace{10pt} r \gg r_\star \,.
\end{array}\right.
\ee
The transition radius $r_\star$ delineating these regimes occurs when $\kappa = m\hat{\mu}$. It can be estimated by substituting the definition of $\kappa$ given in~\eqref{kappadef}, and approximating $\hat{\mu}$ as constant, with value set by the central density as in~\eqref{mualone}: $\hat{\mu}\simeq \rho^2_0/8\Lambda^2m^5$.
The result is
\be
r_\star \simeq \left(\frac{M_{\rm b}}{10^{11}\,M_\odot}\right)^{1/10}\left(\frac{M_{\rm DM}}{M_{\rm b}}\right)^{-2/5} \left(\frac{m}{{\rm eV}}\right)^{-8/5} \left(\frac{\Lambda}{{\rm meV}}\right)^{-8/15} 20 \;{\rm kpc}\,.
\label{rtran}
\ee
For instance, with the fiducial parameters~\eqref{fidparam}, the transition radius for a MW-like galaxy with $M_{\rm b} = 3\times 10^{11}\,M_\odot$ and cosmic DM-baryon ratio $\frac{M_{\rm DM}}{M_{\rm b}} = \frac{\Omega_{\rm DM}}{\Omega_{\rm b}} \simeq 6$ is $r_\star \simeq 49\;{\rm kpc}$. 

\begin{figure}[t]
\centering
\includegraphics[width=6.5in]{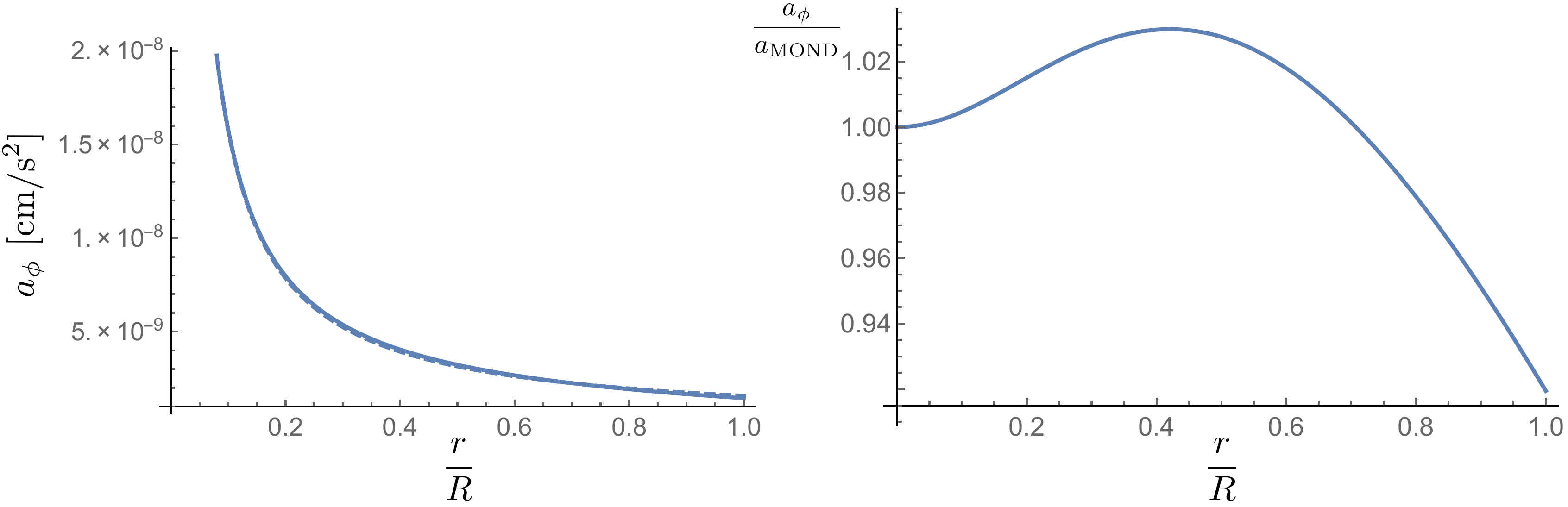}
\caption{\label{aphiMOND} \small {\it Left Panel:} The $\phi$-mediated acceleration $a_\phi$ (solid curve) is compared to the deep-MOND acceleration $a_{\rm MOND}$ (dashed curve) for a MW-like galaxy ($M_{\rm b} = 3\times 10^{11}\,M_\odot$) with cosmological DM-to-baryon ratio $\frac{M_{\rm DM}}{M_{\rm b}} = \frac{\Omega_{\rm DM}}{\Omega_{\rm b}} \simeq 6$, and fiducial values $m = 0.6$~eV and $\Lambda = 0.2~{\rm meV}$. {\it Right Panel:} The ratio of the two accelerations as a function of radius is less than a few percent.}
\end{figure}

Figure~\ref{aphiMOND} plots the numerical solution for $\phi'$, assuming $\beta =2$ and the parameter values listed above. The Left Panel compares the scalar acceleration $a_\phi$ (solid curve) to the MOND acceleration $a_{\rm MOND}$ (dashed curve) as a function of $r$. The Right Panel shows the two accelerations only differing by a few percent, hence the predicted rotation curves are nearly identical to those of MOND. In particular, the ``asymptotic" velocity is indistinguishable from that predicted by MOND (especially taking into account the uncertainties in the mass-to-light ratio), and the BTFR follows identically.

It remains to compare the scalar acceleration $a_\phi$ to the Newtonian acceleration $a_{\rm DM}$ due to the DM condensate profile. As we are about to show,
in the MOND regime ($r\ll r_\star$) the gravitational acceleration from the DM halo is negligible compared to the scalar-mediated MOND acceleration. In the opposite regime ($r \gg r_\star$), on the other hand, the DM halo gives the dominant contribution to the force on a test baryonic particle. 

First, consider the regime $r\ll r_\star$ where $a_\phi$ is approximately MONDian. In this case we have $\phi'^2\simeq \kappa \gg m\hat{\mu}$, hence the DM density profile is
\bea
\nonumber
\rho_{\rm DM} & = & (2m)^{3/2} m  \Lambda \sqrt{\beta-1}  \sqrt{|X|} \\
& \simeq &  2m^2\Lambda \sqrt{\beta-1} \sqrt{\kappa(r)}\,,
\eea
where we have made use of the substitution $\Lambda\rightarrow \sqrt{\beta-1}  \Lambda$ mentioned earlier. Thus $\rho_{\rm DM}\sim 1/r$, and the Poisson equation
can be straightforwardly integrated (ignoring baryons) to obtain $a_{\rm DM} = \frac{m^2 \Lambda \sqrt{\beta-1}}{2M_{\rm Pl}^2} \sqrt{\kappa(r)r^2}$, 
which is constant. Comparing to the scalar acceleration $a_\phi \simeq \frac{\alpha\Lambda}{M_{\rm Pl}} \sqrt{\kappa(r)}$, we find
\be
\frac{a_{\rm DM}}{a_\phi} = \frac{\sqrt{\beta-1}}{2\alpha} \frac{m^2r}{M_{\rm Pl}} \sim 0.6 \;\frac{r}{r_\star}\; \qquad (r\ll r_\star)\,,
\label{ratiosmallr}
\ee
where in the last step we have assumed the parameter values listed below~\eqref{rtran} for concreteness. Hence, as claimed, the gravitational acceleration due to the DM halo
is subdominant in the MONDian regime ($r\ll r_\star$) and becomes comparable to the scalar-mediated acceleration around the transition radius~$r\sim r_\star$. 

Consider now the opposite regime, $r \gg r_\star$. In this case we have $X\simeq \hat{\mu} $, and the DM halo approximates the Lane-Emden density profile found in Sec.~\ref{haloprof}. The gravitational acceleration is $a_{\rm DM} = \frac{1}{m}|X'| \sim \frac{X}{mR}$, while the scalar acceleration is $a_\phi \simeq \frac{\alpha\Lambda}{M_{\rm Pl}} \phi'$,
with $\phi'\simeq   \sqrt{\frac{3}{2m\hat{\mu}}}\sqrt{\frac{\beta-1}{2\beta-3}} \kappa$. For the parameter values listed below~\eqref{rtran} and taking $\beta = 2$ for concreteness, their ratio is given by
\be
\frac{a_{\rm DM}}{a_\phi}  \sim \left(\frac{r}{r_\star}\right)^2 \; \qquad (r\gg r_\star)\,.
\label{ratiolarger}
\ee
Despite the crudeness of the estimate, this is remarkably consistent with~\eqref{ratiosmallr} for $r \sim r_\star$. 
Hence, for $r \gg r_\star$, the DM halo gives the dominant contribution to the acceleration on a test baryonic particle, as claimed earlier.

Let us check the stability of the phonon background. Expanding~\eqref{LphonT} to quadratic order in perturbations $\varphi = \phi - \bar{\phi}(r)$,
we obtain
\bea
\nonumber
{\cal L}_{\rm quad} &=&  \frac{\Lambda(2m)^{1/2}}{\bar{Z}^{3/2}} \Bigg\{ \left[(\beta-1)\hat{\mu} +\left(\frac{\beta}{3} + 1\right) \frac{\bar{\phi}'^2}{2m} \right] \frac{m(\beta-1)\dot{\varphi}^2}{2}  \\
\nonumber
& &~~~~~~~~~~~~~ - \left[ (\beta-1)\left(\frac{2\beta}{3}- 1\right)\hat{\mu} + \left(\frac{\beta}{3}- 1\right)  \frac{\bar{\phi}'^2}{2m} \right] \bar{\phi}' \dot{\varphi}\varphi'  \\
\nonumber
& &~~~~~~~~~~~~~ - \left[(\beta-1)\left(\frac{2\beta}{3}- 1\right)\hat{\mu}^2 + \frac{3\bar{\phi}'^2}{2m}(\beta-1)\hat{\mu} +  \frac{\bar{\phi}'^4}{2m^2}\right]\varphi'^2 \\
& &~~~~~~~~~~~~~ - \left[\left(\frac{2\beta}{3}- 1\right)\hat{\mu} + \frac{\bar{\phi}'^2}{2m}\right]\frac{\bar{Z}(\partial_\Omega\varphi)^2}{r^2}\Bigg\} \,,
\eea
where $\bar{Z} \equiv (\beta-1) \hat{\mu}  + \frac{\bar{\phi}'^2}{2m}$. The sign of the kinetic term is healthy if $\beta > 1$. 
Moreover, the sign of the $(\partial_\Omega\varphi)^2$ term is correct if $\beta \geq 3/2$, which ensures there are no gradient instabilities along the
angular directions. Along the radial direction, the sign of the $\varphi'^2$ is also correct if $\beta \geq 3/2$. It is then trivial to check
by diagonalizing the kinetic matrix that radial perturbations propagate with the correct signature, {\it i.e.}, they are free of ghosts or gradient instabilities.
To summarize, the phonon background is perturbatively stable if $\beta \geq 3/2$, as claimed earlier. 

Note that, in the MOND regime ($\bar{\phi}'^2 \gg 2m\hat{\mu}$), the phonon sound speed is $c_{\rm s} \sim \bar{\phi'}/m$, which is enhanced compared to the sound speed~\eqref{cs} $c_s = \sqrt{2\mu/m}$ computed in the absence of baryons. When we discuss various astrophysical probes below, we will nevertheless apply Landau's criterion for the onset of dissipative effects using $c_s = \sqrt{2\mu/m}$, keeping in mind that this is conservative (since the actual sound speed is in fact larger).

\section{Validity of EFT and the Solar System}

Our background solution~\eqref{phigensoln} involves large phonon gradients, $\frac{\phi'^2}{2m}\gg \mu$, so naturally one should wonder whether it lies within
the regime of validity of the EFT. First notice that in terms of the superfluid velocity $v_{\rm s} \equiv |\vec{\nabla}\phi|/m$ and sound speed $c_s = \sqrt{2\mu/m}$,
the MONDian regime $\frac{\phi'^2}{2m}\gg \mu$ precisely corresponds to $v_{\rm s} \gg c_s$. It therefore violates Landau's criterion $v_{\rm s} \;\lsim\; c_s$ for the stability of superfluid flow. This is of course not surprising --- Landau's criterion is based on the stability of the superfluid against the creation of collective excitations, whereas we wish to work in a regime where baryons generate a large coherent
phonon background. 

As a check on whether this is legitimate, we can compare higher-derivative corrections to the leading-order $P(X)$ Lagrangian. Such corrections by definition involve
more than one derivative per field, hence they can be neglected as long as
\be
\frac{\phi''}{\Lambda_{\rm s}\phi'} \sim \frac{1}{\Lambda_{\rm s}r} \ll 1\,.
\ee
But since $\Lambda_{\rm s}\sim {\rm meV}$, as we have seen in~\eqref{Lsexplicit}, this condition is trivially satisfied on astrophysical scales of interest.
In other words, the phonon profile generated by a galaxy, while large relative to $\mu$, is nevertheless very smooth on the scale of the cutoff.

On the other hand, we should also verify that the local superfluid velocity does not exceed the BEC critical velocity,
\be
v_{\rm s} \ll  v_{\rm c} \sim \left(\frac{\rho}{m^4}\right)^{1/3}\,,
\label{vsvc}
\ee
for otherwise the large phonon gradient will induce a local loss of coherence of the condensate. Equivalently,~\eqref{vsvc} can be understood as the requirement
that the superfluid de Broglie wavelength $\lambda \sim \frac{1}{mv_{\rm s}}$ is much larger than the interparticle separation $\ell\sim \left(\frac{m}{\rho}\right)^{1/3}$. To estimate $v_{\rm c}$, we use the
halo mass density $\rho = (2m)^{3/2}m\Lambda\sqrt{|X|} \simeq 2m^2\Lambda\sqrt{\kappa}$, where in the last step we have assumed the MOND regime.
This gives
\be
v_{\rm c} \simeq 0.025 \left(\frac{M_{\rm b}}{10^{11}\,M_\odot}\right)^{1/6} \left(\frac{m}{{\rm eV}}\right)^{-2/3} \left(\frac{\Lambda}{{\rm meV}}\right)^{2/9}\left(\frac{{\rm kpc}}{r}\right)^{1/3} \,.
\label{vclocal}
\ee
Meanwhile, the superfluid velocity is $v_{\rm s} = \phi'/m \simeq \sqrt{\kappa}/m$, which gives
\be
v_{\rm s} \simeq 0.008  \left(\frac{M_{\rm b}}{10^{11}\,M_\odot}\right)^{1/2} \left(\frac{m}{{\rm eV}}\right)^{-1} \left(\frac{\Lambda}{{\rm meV}}\right)^{-1/3}\; \frac{{\rm kpc}}{r}\,.
\label{vslocal}
\ee
Thus the criterion~\eqref{vsvc} can be expressed as a bound on the distance from the galactic center:
\be
r \gg 0.2  \left(\frac{M_{\rm b}}{10^{11}\,M_\odot}\right)^{1/2} \left(\frac{m}{{\rm eV}}\right)^{-1/2} \left(\frac{\Lambda}{{\rm meV}}\right)^{-5/6}\; {\rm kpc}\,.
\ee
This is satisfied down to the central regions of galaxies.

The condition~\eqref{vsvc} does have important ramifications for the solar system. It is well-known within the standard MOND framework that the extra acceleration $a_\phi$, albeit small compared to the Newtonian acceleration in the solar system, gives an unacceptably large correction to Newtonian gravity, in conflict with bounds from tests of gravity. One possible way out is to suitably modify $P(X)$ at large $X$, but this requires fine-tuning~\cite{Bruneton:2007si}. Another possibility is to introduce a suitable higher-derivative galileon operator~\cite{Babichev:2011kq}, but this has the obvious disadvantage of complicating the theory. 

In our superfluid picture, we are naturally immune to this problem because the local phonon gradient generated by the Sun is so large that~\eqref{vsvc} is violated throughout the solar system.
Indeed, the superfluid velocity~\eqref{vslocal} due to the Sun ($M_{\rm b} = M_\odot$) is
\be
v_{\rm s}^\odot \simeq 5 \left(\frac{m}{{\rm eV}}\right)^{-1} \left(\frac{\Lambda}{{\rm meV}}\right)^{-1/3}\; \frac{{\rm AU}}{r}\,,
\ee
where $r$ now represents the distance from the Sun. Meanwhile the BEC critical velocity~\eqref{vclocal} is set by the Milky Way galaxy ($M_{\rm b} = 3\times 10^{11}\;M_\odot$) evaluated at the location of the solar system ($\sim 8\;{\rm kpc}$ from the galactic center):
\be
v_{\rm c}^{\rm MW} \simeq 0.02 \left(\frac{m}{{\rm eV}}\right)^{-2/3} \left(\frac{\Lambda}{{\rm meV}}\right)^{2/9}  \,.
\ee
The criterion~\eqref{vsvc} can be expressed as a bound on the distance from the Sun:
\be
r \gg 250 \left(\frac{m}{{\rm eV}}\right)^{-1/3} \left(\frac{\Lambda}{{\rm meV}}\right)^{-5/9}\; {\rm AU}\,, 
\ee
which is larger than the solar system.

The fact that~\eqref{vsvc} is violated in the solar system means that the BEC loses its coherence, and the condensate is replaced by a phase of normal DM. 
Hence the usual worries about MOND and local tests of gravity do not apply in our case. Furthermore, since our DM behaves as ordinary particles in
the solar system, this is good news for direct detection experiments. By allowing the usual axion-like couplings to Standard Model operators,
our DM particles can be detected through the suite of standard axion-like particle searches, {\it e.g.},~\cite{Essig:2013lka}.

\section{A Relativistic Completion}
\label{phi6}

It is well-known that a superfluid can be described in the weak-coupling regime as a theory of a self-interacting complex scalar field with global $U(1)$ symmetry.
The conserved charge associated with this symmetry is the total number of particles. A superfluid corresponds to a state which spontaneously breaks the
global $U(1)$ and has finite charge density under this symmetry. 

In this Section we give an explicit example of such a theory that admits a condensate with $P\sim\mu^{3/2}$ equation of state. After integrating out the radial
mode, the resulting action for the phase to leading order in derivatives will be exactly given by~\eqref{Lphon}, with the desired square root. The first theory that 
comes to mind is a scalar with hexic interactions, $\mathcal{L}=-|\partial_\mu \Phi |^2-m^2 |\Phi |^2- \lambda|\Phi|^6$.
As shown in the Appendix, this gives $P(X) \sim X^{3/2}$, exactly the desired fractional power for MOND. However the sign is wrong.
For a stable potential ($\lambda > 0$), one is restricted to $X > 0$, hence spatial gradients can never dominate and the MOND regime is inaccessible.
The MOND regime is only possible for $\lambda < 0$, but this branch is of course unstable.

Instead we will consider the following theory
\be
{\cal L} = -\frac{1}{2} \left(|\partial_\mu \Phi |^2+m^2 |\Phi |^2\right) -  \frac{\Lambda^4}{6\left(\Lambda_{\rm c}^2 + |\Phi |^{2}\right)^6}\left(|\partial_\mu \Phi |^2+ m^2 |\Phi |^2\right)^3\,.
\ee
The scale $\Lambda_{\rm c}$ is introduced to ensure that the theory admits a $\Phi = 0$ vacuum. The MOND regime corresponds to $|\Phi|^2\gg \Lambda_{\rm c}^2$, as we will see shortly. 
Notice the absence of a quartic term $(\left(|\partial_\mu \Phi |^2 + m^2 |\Phi |^2\right)^4$. It is possible to include such a term provided its coefficient is not too large, as we will see towards the end of the Section.

For our purposes it suffices to focus on the non-relativistic regime. Making the field redefinition
\be
\Phi= \rho  e^{i(\theta + mt)}\,,
\label{NRfieldredef}
\ee
and taking the non-relativistic limit, it is straightforward to arrive at
\be
{\cal L} = -\frac{1}{2}\left((\vec{\nabla}\rho)^2 - 2m\rho^2 X\right) - \frac{\Lambda^4}{6\left(\Lambda_{\rm c}^2 + \rho^2\right)^6}\left((\vec{\nabla}\rho)^2 - 2m\rho^2 X\right)^3 \,.
\label{Lwithrho}
\ee
The power of $\rho$ in the denominator of the second term guarantees the MOND scaling symmetries~\cite{Milgrom:1997gx,Milgrom:2008cs}: assuming that spatial gradients dominate, and taking the MOND limit $\rho \gg \Lambda_{\rm c}$, the action is invariant under the spatial scaling
\be
h_{ij} \rightarrow \Omega^2 h_{ij}\,;\qquad \rho \rightarrow \Omega^{-1/2} \rho\,.
\label{MONDsyms}
\ee

The effective theory of the Goldstone mode is obtained by integrating out $\rho$. To leading order in the derivative expansion we can ignore $(\vec{\nabla}\rho)^2$ contributions.
In this limit the equation for $\rho$ becomes algebraic:
\be
2mX\rho\left[\left(\Lambda_{\rm c}^2 + \rho^2\right)^7 +\Lambda^4(2mX)^2\rho^4\left(\Lambda_{\rm c}^2 - \rho^2\right)\right] = 0\,.
\ee
The MOND regime corresponds to $\rho \gg \Lambda_{\rm c}$. Indeed, in this limit 
the solution is
\be
\rho^2 \simeq  \Lambda \sqrt{2m} \left(X^2\right)^{1/4} = \Lambda \sqrt{2m|X|} \,.
\label{rhosoln}
\ee
Substituting this back into~\eqref{Lwithrho} gives, to leading order in derivatives, 
\be
{\cal L} \simeq \frac{2\Lambda(2m)^{3/2}}{3}X\sqrt{|X|} \,.
\label{LMONDfin}
\ee
This agrees with the MOND phonon action~\eqref{Lphon}.

The regulator scale $\Lambda_{\rm c}$ implies that the MOND regime is restricted to $\rho \;\gsim\; \Lambda_{\rm c}$, {\it i.e.},
$|X| \;\gsim \; \frac{\Lambda_{\rm c}^4}{2m\Lambda^2}$. Using~\eqref{aphi}, this corresponds to
\be
a_\phi \;\gsim\;  \frac{\Lambda_{\rm c}}{\alpha^2\Lambda} \; a_0\,.
\ee
Observationally the MOND regime works quite well down to $\sim a_0/10$, so this puts an upper bound on $\Lambda_{\rm c}$. By choosing $\Lambda_{\rm c}$
a factor of a few smaller than $\Lambda$, the predicted breakdown could occur around the acceleration scale of the MW dwarf spheroidals, which are well-known to pose a
challenge for MOND~\cite{Spergel,Milgrom:1995hz,Angus:2008vs,Hernandez:2009by,Lughausen:2014hxa}.

We can straightforwardly generalize the analysis to include a quartic term. To fast-track the discussion, let us immediately write the answer in terms of polar variables:
\be
{\cal L} = -\frac{1}{2}\left((\vec{\nabla}\rho)^2 - 2m\rho^2 X\right)  + \frac{g\Lambda^2}{2\left(\Lambda_{\rm c}^2 + \rho^2\right)^3}\left((\vec{\nabla}\rho)^2 - 2m\rho^2 X\right)^2 - \frac{\Lambda^4}{6\left(\Lambda_{\rm c}^2 + \rho^2\right)^6}\left((\vec{\nabla}\rho)^2 - 2m\rho^2 X\right)^3\,,
\label{Lquartic}
\ee
where $g$ is dimensionless. The power of $\rho$ in the denominator of the new term is once again chosen such that~\eqref{MONDsyms}
is a symmetry when $\rho \gg \Lambda_{\rm c}$ and spatial gradients dominate. Focusing on this regime for simplicity, the
equation of motion for $\rho$ is a quadratic equation for $\rho^4$. Choosing the branch such that the answer reduces to~\eqref{rhosoln} as $g\rightarrow 0$, we find
\be
\rho^2 \simeq \Lambda \sqrt{- g m X   + 2m|X| \left(1 + \frac{g^2}{4}\right)^{1/2} }\,.
\ee
Upon substituting into~\eqref{Lquartic}, the action for the Goldstone will of course be different than~\eqref{LMONDfin}, but will reduce to it in the limit of large $X$.
What matters ultimately is that the Lagrangian for the Goldstone has the same sign as~\eqref{LMONDfin}, for both $X$ positive and negative. It is easy to show
that this is the case for
\be
g^2 < \frac{4}{3}\,.
\ee
Clearly the analysis can be generalized even further by including higher order terms, $(\left(|\partial_\mu \Phi |^2 + m^2 |\Phi |^2\right)^n$, with $n \geq 4$,
provided they respect the scaling symmetry~\eqref{MONDsyms} in the appropriate limit. Their coefficients will be similarly constrained.

\section{Cosmology}
\label{cosmo}

In this Section we study the cosmology of the DM superfluid. As mentioned towards the end of Sec.~\ref{superfluid} the simplest genesis scenario is through
a vacuum displacement mechanism, with DM being generated at a time when $H_i\sim m$ corresponding to a baryon-photon temperature of order 50~TeV. The DM
is initially very cold, it rapidly reaches thermal equilibrium with itself, but is decoupled from ordinary matter to first approximation. 

In order to obtain an acceptable background cosmology and linear perturbation growth, we will see that the $\Lambda$ and $\alpha$ parameters of the phonon EFT must assume different values cosmologically than in galaxies. This is not unreasonable, as argued in Sec.~\ref{superphase}, since these parameters are expected to depend on $T/T_{\rm c}$, and this ratio is $\sim 22$ orders of magnitude smaller cosmologically than in galaxies. Furthermore, we have already invoked finite-temperature effects in galaxies in Sec.~\ref{2fluid} to ensure stability of the MOND regime. We will denote the cosmological values by $\Lambda_0$ and $\alpha_0$.

\subsection{Equation of State}
\label{eoscosmo}

The first thing to check is whether the condensate has sufficiently small pressure to act as dust. Recall from~\eqref{eos} our condensate equation of state:
\be
w = \frac{P}{\rho} = \frac{\rho^2}{12\Lambda^2_0m^6}\,.
\label{w0}
\ee
The sound speed of linear fluctuations is identical, $c_s^2 = w$. At sufficiently low density ($\rho \ll \Lambda_0 m^3$) the superfluid behaves as dust, whereas at high density ($\rho \gg \Lambda_0 m^3$) it behaves as a relativistic component. At the very least, we should impose that $w \ll 1$ at matter-radiation equality. Since  $w \sim 1/a^6$, and correspondingly $c_s \sim 1/a^3$,
imposing $w_{\rm eq} \ll 1$ will ensure that DM behaves to a very good approximation as dust throughout the matter-dominated era. Substituting the known value $\rho_{\rm eq} \simeq 0.4~{\rm eV}^4$, this puts a lower bound on $\Lambda_0$:
\be
\Lambda_0\gg 0.1 \left(\frac{m}{{\rm eV}}\right)^{-3}\;{\rm eV}\,.
\label{L0}
\ee
In particular $\Lambda_0\gg 0.5~{\rm eV}$ for our fiducial value $m = 0.6$~eV. This is roughly four orders of magnitude larger than the fiducial value $\Lambda = 0.2$~eV
assumed in galaxies. This can be achieved, for instance, if $\Lambda$ depends on temperature as
\be
\Lambda(T) =  \frac{\Lambda_0}{1 + \kappa_\Lambda (T/T_{\rm c})^{1/4}}\,;\qquad \kappa_\Lambda \sim 10^4\,.
\ee

\subsection{Coupling to baryons}

The above equation of state was derived ignoring the coupling to baryons. We now rectify this and derive the phonon cosmological evolution sourced by the baryonic density.
Setting $\theta = \theta(t)$, the phonon action given by~\eqref{Lphon} and~\eqref{Lint} becomes
\be
{\cal L} =  \frac{2\Lambda_0 (2m)^{3/2}}{3}a^3\dot{\theta}^{3/2}  - \alpha_0 \frac{\Lambda_0}{M_{\rm Pl}}a^3 \theta \rho_{\rm b}\,.
\ee
Varying with respect to $\theta$ gives the equation of motion
\be
\frac{{\rm d}}{{\rm d}t} \left((2m)^{3/2}a^3\dot{\theta}^{1/2}\right) =  -\frac{\alpha_0}{M_{\rm Pl}}a^3\rho_{\rm b}\,.
\ee
Since $a^3\rho_{\rm b} = {\rm const.}$, we can integrate straightforwardly: $(2m)^{3/2}\dot{\theta}^{1/2} = -\frac{\alpha_0}{M_{\rm Pl}}\rho_{\rm b} t  + \frac{C}{a^3}$,
where $C$ is an integration constant. In the non-relativistic approximation, the energy density is $\rho = m n = m\Lambda_0 (2m)^{3/2}\dot{\theta}^{1/2}$,
hence
\be
\rho =  -\frac{\alpha_0\Lambda_0}{M_{\rm Pl}}mt \rho_{\rm b}  +   \rho_{\rm dust}\,,
\ee
where $\rho_{\rm dust} = m\Lambda_0 C/a^3$. This term is recognized as the dust contribution studied in the
(baryon-free) analysis of Sec.~\ref{eoscosmo}. Note that the non-relativistic approximation breaks down when
$\dot{\theta} \sim m$, corresponding to $\rho \sim m^3\Lambda_0$, which from~\eqref{w0} is precisely when pressure becomes non-negligible.

In the matter-dominated era, $t \sim a^{3/2}$, the baryonic contribution $\sim \rho_{\rm b} t$ redshifts as $1/a^{3/2}$ whereas the second term redshifts as usual as $\rho_{\rm dust}\sim 1/a^3$. In order for the superfluid to behave as ordinary dust, the second term should dominate over the first all the way to the present time:
\be
\frac{\alpha_0\Lambda_0}{M_{\rm Pl}}mt_0 \frac{ \rho_{\rm b}}{\rho_{\rm dust}} \;\lsim\; 1\,.
\ee
Substituting the age of the universe $t_0 = 13.9\times 10^9~{\rm yrs} \simeq 6 \times 10^{32}~{\rm eV}^{-1}$, 
and assuming a DM-to-baryon ratio of $\rho_{\rm dust}/\rho_{\rm b} = 6$, we obtain
\be
\alpha_0 \;\lsim\; 2.4 \times 10^{-5} \frac{{\rm eV}^2}{\Lambda_0m} \ll  2.4 \times 10^{-4 }\left(\frac{m}{{\rm eV}}\right)^2\,,
\label{al0}
\ee
where the last step follows from~\eqref{L0}. In particular, $\alpha_0 \ll 10^{-4}$ for our fiducial value $m = 0.6$~eV.
This is roughly four orders of magnitude smaller than the value $\alpha = 2.5$ obtained in galaxies by matching to MOND.
This can be achieved, for instance, if $\alpha$ depends on temperature as
\be
\alpha(T) =  \alpha_0\left(1 + \kappa_\alpha (T/T_{\rm c})^{1/4}\right)\,;\qquad \kappa_\alpha \sim 10^4\,.
\label{aIT}
\ee

Note that while $\Lambda(T)$ and $\alpha(T)$ both depend on temperature, the scale $\Lambda' \sim \alpha \Lambda$ appearing in the phonon-baryon
coupling~\eqref{Lint} is nearly temperature-independent. 

\subsection{Velocity-dependent critical acceleration}

An immediate corollary of $\Lambda(T)$ and $\alpha(T)$ being temperature-dependent is that the critical acceleration,
\be
a_0 \sim \alpha \frac{(\alpha \Lambda)^2}{M_{\rm Pl}}\,,
\ee
also depends on temperature. More precisely, since the product $\alpha\Lambda$ is constant to a first approximation, the temperature dependence of $a_0$
is governed by $\alpha$. In particular, in light of~\eqref{al0} we obtain
\be
a_0^{\rm cosmo}  \ll 10^{-4} a_0\,,
\ee
where $a_0$ is the typical MOND value~\eqref{critacc} in galaxies. Given this strong suppression of $a_0$, it follows that gravity is highly Newtonian on cosmological scales.

Another consequence is that there is no longer a universal value for the MOND critical acceleration in galaxies, instead $a_0$ is predicted to depend on the velocity dispersion. The functional dependence is model-dependent of course, but the generic trend is that $a_0$ {\it decreases with decreasing velocity}. Intriguingly, this trend has been noted in the data --- low-surface brightness galaxies tend to prefer a lower value of $a_0$~\cite{Swaters:2010qe}.

\section{Gravitational Lensing}
\label{lensing}

In the context of TeVeS~\cite{Bekenstein:2004ne}, the absence of DM in galaxies forces one to assume a rather complicated coupling between the scalar field $\phi$ and matter fields in order to reproduce acceptable gravitational lensing. For starters, one supplements the theory with a 4-vector field $A_\mu$, which is unit time-like $g^{\mu\nu} A_{\mu} A_\nu = -1$. Then the non-relativistic scalar-matter interaction ${\cal L}_{\rm coupling} = -\frac{\alpha\Lambda}{M_{\rm Pl}} \phi \rho_{\rm b}$ is covariantized by coupling matter fields to an effective metric $g_{\mu\nu}^{\rm TVS}$, defined in terms of the Einstein-frame metric $g_{\mu\nu}$ via
\bea
\nonumber
g_{\mu\nu}^{\rm TVS} &=& e^{-\frac{2\alpha\Lambda}{M_{\rm Pl}} \phi } g_{\mu\nu} - 2A_\mu A_\nu \sinh \frac{2\alpha\Lambda}{M_{\rm Pl}} \phi \\
&\simeq & g_{\mu\nu} - \frac{2\alpha\Lambda}{M_{\rm Pl}} \phi \Big(g_{\mu\nu} +2 A_\mu A_\nu\Big) \,.
\label{gTVS}
\eea
In the weak-field, quasi-static regime, $g_{\mu\nu}$ takes the usual form: $g_{00} = -(1 + 2\Phi)$, $g_{0i} = 0$ and $g_{ij} = (1-2\Phi)\delta_{ij}$.
To this order we can ignore perturbations in the vector field, {\it i.e.}, $A_\mu = (1,0,0,0)$, such that
\be
{\rm d}s^2_{\rm TVS}  \simeq -\left(1 + 2\left[\Phi + \frac{\alpha\Lambda}{M_{\rm Pl}} \phi \right] \right){\rm d}t^2 + \left(1 - 2\left[\Phi  + \frac{\alpha\Lambda}{M_{\rm Pl}} \phi  \right]\right){\rm d}\vec{x}^2\,,
\label{lensingform}
\ee
where $\Phi$ is of course sourced by baryons only:
\be
\nabla^2\Phi = 4\pi G_{\rm N} \rho_{\rm b} \,.
\ee
This line element is exactly of the General Relativity form, albeit in terms of a shifted gravitational potential $\Phi +  \frac{\alpha\Lambda}{M_{\rm Pl}}\phi$.
Hence the mass inferred from lensing observations matches the mass inferred from dynamical measurements.
The TeVeS metric~\eqref{gTVS} was of course precisely engineered for this purpose. Specifically, the equality of gravitational potentials in~\eqref{lensingform}
traces back to the precise factor of 2 in the combination $g_{\mu\nu} +2 A_\mu A_\nu$ appearing in~\eqref{gTVS}. This relative factor is not protected by any symmetry.

In our case the story is simpler on two counts. First, there is no need to postulate an additional vector field. The normal fluid
component already provides us with a time-like vector field $u^\mu$, as discussed in Sec.~\ref{2fluid}.
Second, the DM in galaxies contributes to lensing, hence the TeVeS factor of 2 can be generalized:
\be
\tilde{g}_{\mu\nu} \simeq g_{\mu\nu} - \frac{2\alpha\Lambda}{M_{\rm Pl}} \phi \Big(\gamma g_{\mu\nu} +(1 + \gamma) u_\mu u_\nu\Big)\,,
\label{ourg}
\ee
with $\gamma= 1$ corresponding to the TeVeS tuning. Working in the rest frame of the normal fluid, this gives in the weak-field limit,
\be
{\rm d}\tilde{s}^2  \simeq -\left(1 + 2\left[\Phi + \frac{\alpha\Lambda}{M_{\rm Pl}} \phi \right] \right){\rm d}t^2 + \left(1 - 2\left[\Phi  + \gamma \frac{\alpha\Lambda}{M_{\rm Pl}} \phi  \right]\right){\rm d}\vec{x}^2\,,
\label{ourlensingform}
\ee
where $\Phi$ is now sourced by both baryonic and dark matter:
\be
\nabla^2\Phi = 4\pi G_{\rm N} \left(\rho_{\rm b} + \rho_{\rm DM}\right)\,.
\ee
Hence the lensing signal will arise from a combination of the $\gamma$ term in~\eqref{ourlensingform} and the DM condensate density profile
shown in Fig.~\ref{laneemden}. Determining the allowed range of $\gamma$ will require a detailed comparison with lensing observations, which is beyond the scope of this paper.
What is clear is that there should be considerably more freedom than in TeVeS. It may even be
that $\gamma = -1$ is allowed, in which case the coupling to matter would reduce to a simple conformal coupling.

In most of our discussion so far, we have assumed fiducial parameter values~\eqref{fidparam} such that the condensate radius
is of order the virial radius, {\it e.g.} $R\sim 158$~kpc for $M_{\rm DM} = 10^{12}\,M_\odot$  compared to 200~kpc for the virial radius.
By choosing other parameter values, however, we can consider smaller condensate radii, in which case the condensate core will be surrounded by an envelope of DM particles in the normal phase, presumably with a NFW density profile. In that case the lensing signal could result primarily from the NFW envelope. This deserves a dedicated analysis, which will
appear elsewhere. 

\section{Merging Clusters: the Bullet and the Counter-Bullet}
\label{bullet}

The ``Bullet'' Cluster 1E0657-57~\cite{Clowe:2003tk,Clowe:2006eq,Bradac:2006er} shows lensing peaks displaced from the gas and centered around the galaxy distribution.
This is expected in CDM: the halos are made up of weakly interacting dark matter particles that fly past each other, together with the galaxies, while the
baryonic plasma is slowed down by ram pressure and ends up spatially segregated from the halos. By now observers have identified over thirty such merging
systems~\cite{Diego:2014eda,Harvey:2015hha}. 
 
Galaxy clusters in the present context are comprised, either partially or fully, of DM particles in the normal phase. Hence we also expect lensing peaks displaced from the gas,
due to the DM component. An important consideration is the constraint this imposes on the self-interaction cross section of the DM~\cite{Randall:2007ph,Massey:2010nd}.
The tightest constraint comes a recent analysis of $\sim 30$ merging systems~\cite{Harvey:2015hha}:
\be
\frac{\sigma}{m} \;\lsim\;  0.5~\frac{{\rm cm}^2}{\rm g} \,.
\label{bulletcons}
\ee
At face value there is a window for which this is consistent with our lower bound~\eqref{siglow} for DM condensation in galaxies. 
However we think that the constraint~\eqref{bulletcons} is not as stringent in our case. Indeed,~\eqref{bulletcons} was derived assuming a single DM component, whereas the 2-fluid mixture makes for a much richer situation. The heterogeneous nature of merging systems, with different interactions among their components, can result in a significantly weaker bound in our case.\footnote{This loophole was also exploited recently with ultra-strongly interacting DM~\cite{Pollack:2014rja}.} Specifically, we expect the superfluid components to pass through each other with negligible dissipation if the relative velocity is sub-sonic, 
\be
v_{\rm infall} \; \lsim\; c_s\,.
\ee
Using~\eqref{cs} and~\eqref{mur}, and assuming the fiducial parameter values~\eqref{fidparam} for concretenes, it is straightforward to show that 
$c_s \simeq 1400~{\rm km}/{\rm s}$ for the sub-cluster ($M_{\rm sub} \simeq 10^{14}M_\odot$), while $c_s \simeq 3500~{\rm km}/{\rm s}$ for the main cluster ($M_{\rm main} \simeq 10^{15}M_\odot$), assuming a significant fraction of their mass is condensed. These values are comparable to the estimate of $\sim 2700$~km/s for the relative velocity~\cite{Springel:2007tu,lagefarrar}, indicating that dissipative processes between the superfluid cores should be suppressed.\footnote{This is unlike BEC DM, where the critical velocity is only $\simeq 100$~km/s~\cite{slepiangoodman}.}

In general our framework predicts two distinct features that should appear simultaneously in the lensing maps of bullet-like merging systems: $i)$ mass peaks coincident with the cluster galaxies, due to the (non-interacting) superfluid cores; $ii)$ another mass peak, approximately coincident with the X-ray luminosity peak, due to the (interacting) normal components. 
Interestingly, this is consistent with the complex mass structure of the ``train wreck" Abell 520 (MS0451+02) merging system~\cite{Mahdavi:2007yp,Jee:2012sr,Clowe:2012am,Jee:2014hja}, often hailed as a counterexample to the ``Bullet" cluster. Aside from the ``bullet-like" lensing peaks around bright galaxies segregated from the gas, this system also exhibits a puzzling ``dark core" overlapping the X-ray gas without corresponding bright galaxies. In the context of SIDM, the cross section required to explain this feature is inconsistent with the bullet bound~\eqref{bulletcons}~\cite{Jee:2014hja}. In our case, however, the dark core is naturally explained as due the normal DM components.\footnote{It has been argued that the contradictory nature of the Bullet and counter-Bullet can also be explained in the BEC DM context~\cite{Lee:2008mq}.}
Intriguingly, even in the case of the Bullet Cluster the combined strong and weak lensing map reveals a significant mass peak coincident with the X-ray gas~\cite{Bradac:2006er}.

Another way that~\eqref{siglow} and~\eqref{bulletcons} can be satisfied simultaneously is if the cross section is velocity-dependent. 
This is in fact expected for dark atoms, since the cross section between ordinary atoms is generally a rich function of velocity~\cite{Cline:2013pca} due to various
atomic resonances. Such velocity dependence may imply a suppressed cross section in clusters, where the typical virial velocity is $\sim 10$ times larger
than in galaxies. A velocity-dependent cross section was proposed in the SIDM context to simulateously match the inferred profiles of dwarf galaxies
and galaxy clusters~\cite{Firmani:2000ce,Colin:2002nk,Feng:2009mn,Loeb:2010gj,Vogelsberger:2012ku}. 

\section{Vortices}
\label{vortices}

As is well-known, a superfluid cannot rotate uniformly. When spun faster than a critical angular velocity the superfluid develops quantum vortices that carry the angular momentum~\cite{landaubookpart2}. In the context of BEC dark matter, vortex formation was initially considered in~\cite{Silverman:2002qx} and studied in detail subsequently in~\cite{RindlerDaller:2011kx}. For the purpose of this paper,
we shall content ourselves with simple dimensional analysis along the lines of~\cite{Silverman:2002qx}.

We can immediately convince ourselves that our halos rotate much faster than critical velocity. The critical angular velocity for vortex formation in a vessel of radius $R$ is, up to a logarithm factor~\cite{landaubookpart2},
\be
\omega_{\rm cr} \sim \frac{1}{mR^2} \sim 10^{-41} {\rm s}^{-1} \,,
\ee
where we have assumed a halo radius $R\sim 100$~kpc and mass $m\sim {\rm eV}$. On the other hand, the angular frequency of a DM halo of density $\rho$ is
$\omega \sim \lambda\sqrt{G_{\rm N}\rho}$, where $\lambda \equiv \frac{LE^{1/2}}{G_{\rm N}M^{5/2}}$ is the so-called spin parameter, while $L$ and $E$ are the total angular momentum and energy of the halo respectively. From CDM simulations one finds $0.01 \;\lsim\; \lambda \;\lsim\; 0.1$. Substituting a typical density of order $\rho\sim 10^{-25}~{\rm g}/{\rm cm}^3$, we find
\be
\omega\sim 10^{-18}\lambda \; {\rm s}^{-1}\,.
\ee
Hence $\omega \gg \omega_{\rm cr}$, and vortex formation is unavoidable. 

The line density of vortices can also be readily estimated,
\be
\sigma_{\rm v} \sim m\omega \sim 10^{2}\lambda~{\rm AU}^{-2}\,.
\ee
In a galactic halo of radius $R\sim 100$~{\rm kpc}, this means $N_{\rm v}\sim 10^{23}$ vortices in total.
Their core radius is of order the healing length $\xi$, which is estimated as
\be
\xi \sim \frac{1}{mc_s} \sim {\rm mm}\,,
\ee
where we have assumed a halo of mass $M\sim 10^{12}M_\odot$ and used the fiducial parameters~\eqref{fidparam}.
Thus the core radius is an order of magnitude or so larger than the average interparticle separation in galaxies. 

It would be interesting to study whether these vortices can be detected observationally, for instance through gravitational lensing.
This may prove challenging, since their kinetic energy per unit volume is tiny: $\Delta\rho \sim \frac{\omega}{m}\rho \sim 10^{-33}\lambda \rho$.
Substructure lensing may soon be possible with the Atacama Large Millimeter Array~\cite{Hezaveh:2012ai}.

\section{Other Astrophysical Consequences}
\label{othercons}

In this Section we speculate on various astrophysical implications of superfluid DM. For the purpose of this initial paper our discussion
will be quite qualitative, leaving a more careful analysis to the future.\\

\begin{itemize}

\item {\bf Galaxy mergers:} A very interesting question is what happens during galaxy mergers. Following Landau's criterion for superfluidity, the merger dynamics depend on the infall velocity $v_{\rm infall}$ compared to the phonon sound speed $c_s$ within halos. The sound speed in a given halo is generally of order of the virial velocity. For instance, for our fiducial parameter values~\eqref{fidparam} we find $c_s \simeq 220$~km/s in a $10^{12}M_\odot$ halo. If the infall velocity is ultra-sonic, $v_{\rm infall} \;\gsim \; c_s$, the encounter will drive halos out of equilibrium, exciting DM particles out of the condensate. As in $\Lambda$CDM, dynamical friction will lead to a rapid halo merger, and after some time the merged halo will thermalize and condense back to the superfluid state. If the infall velocity is sub-sonic $v_{\rm infall} \;\lsim \; c_s$, on the other hand, the merger time scale will be much longer and involve multiple encounters, as dynamical friction between the superfluid halos will be negligible. This is similar to what happens in MOND~\cite{Nipoti:2007ik,Combes:2009ab}. \\

\item {\bf Reduced dynamical friction:} The overall reduction in dynamical friction due to the superfluid nature of the DM halo alleviates a number of minor problems
with CDM. Instead of being slowed down by dynamical friction, galactic bars in spiral galaxies should achieve a nearly constant velocity, as favored by observations~\cite{Debattista:1997bi}.
This effect has been pointed out in BEC DM~\cite{Goodman:2000tg,Chandra} and MOND~\cite{Combes:2009ab}. Reduced dynamical friction would also help with the M81 group of galaxies --- see~\cite{Kroupa:2014gta} and references therein.

Another interesting system is the Fornax dwarf spheroidal.\footnote{We thank Lam Hui for pointing this out to us.} Five satellite globular clusters
orbit Fornax close enough that they should lie within their host's DM halo, assuming an NFW profile. If so, however, dynamical friction should have caused the globular clusters to rapidly fall towards the center of Fornax~\cite{Fornaxold,Tremaine}. In reality Fornax shows no sign of such mergers. A possible explanation in $\Lambda$CDM is that the Fornax's DM halo is cored, with the globular clusters orbiting on the periphery~\cite{Cole}. In our case, the situation is unclear, due to two competing effects. On the one hand, dynamical friction within Fornax's superfluid DM halo should be reduced, as already mentioned. On the other hand, dynamical friction with stars is enhanced in MOND, thereby reducing the merger time~\cite{SanchezSalcedo:2006fa}. This will require a detailed study.\\

\item {\bf Dark-bright solitons:} Given the large coherence length of the BEC, galaxies in the process of merging should exhibit interference patterns (so-called dark-bright solitons) that have been observed in counterflowing BECs at super-critical velocities, {\it e.g.},~\cite{exptpaper}. This effect has been studied to some extent in ultra-light BEC DM~\cite{Gonzalez:2011yg}. It would be interesting to estimate the spatial extent and lifetime of the fringes to see whether they are potentially observable. It is intriguing to speculate that this can offer an alternative mechanism to generate the spectacular shells seen around elliptical galaxies~\cite{shells}.\footnote{We thank Ravi Sheth for suggesting this idea to us.}\\

\item {\bf Vast planar structures and tidal dwarfs:} The vast planar structures seen in the Local Group~\cite{Kroupa:2004pt,Pawlowski:2012vz,Pawlowski:2013kpa,Pawlowski:2013cae,Ibata:2013rh,Conn:2013iu,Ibata:2014pja} and beyond~\cite{Ibata:2014csa}  find a possible explanation in our scenario, similar to that proposed in MOND~\cite{Pawlowski:2012vz}. Namely, the planar structures around the MW and Andromeda would be the result of tidal stripping during a fly-by encounter between these galaxies. In particular, most of their satellite galaxies would be tidal dwarfs. With the MOND force law it has been estimated that MW and Andromeda had a fly-by encounter $\sim 10$~Gyr ago, with~$\lsim \; 55$~kpc closest approach distance~\cite{Zhao:2013uya}. In $\Lambda$CDM, such a past encounter, while in principle possible, would have disastrous consequences: dynamical friction between the extended halos would cause a rapid merger of MW and M31. 
In MOND, however, there is only stellar dynamical friction and a merger can be avoided~\cite{Nipoti:2007ik,Tiret:2007fy,Combes:2009ab}. 

Similarly, in our case dynamical friction is suppressed among DM particles if the infall velocity is sub-sonic, as mentioned before. If even a tiny amount of superfluid DM is stripped along with the tidal dwarf galaxies created in the process, their dynamics will be governed by MOND, resulting in flat rotation curves that fall on the BTFR, consistent with observations of the NGC5291 dwarfs~\cite{Bournaud:2007sz,Gentile:2007gp}. \\

\item {\bf Globular clusters:} It is well-known that globular clusters contain negligible amount of~DM. Indeed, their observations are well-fitted by 
taking only the baryonic mass into account and assuming Newtonian gravity. This poses a problem for MOND~\cite{Ibata:2011ri}. Our case is clearly different, since
the presence of a significant DM component is necessary for the MOND phenomenon to occur. To the extent that whatever mechanism ({\it e.g.}, tidal stripping)
responsible for DM removal in $\Lambda$CDM is also effective in our case, our model predicts DM-free (and therefore MOND-free) globular cluster dynamics.\\

\item {\bf Tri-axial DM halos:} A key prediction of collisionless CDM simulations is the ellipticity of~DM halos~\cite{Dubinski:1991bm}, which is borne out by lensing observations. 
Lensing mass reconstruction of galaxy clusters often require an elliptical DM clump around the brightest central galaxy. On the other hand, DM self-interactions tend to isotropize the DM distribution, resulting in more spherical halos. To match the ellipticity of galaxy cluster MS2137-23 inferred from strong lensing observations,~\cite{MiraldaEscude:2000qt} claimed an even tighter bound than~\eqref{bulletcons}, though recent SIDM simulations find consistent halo morphology for cross sections as large as $\sim {\rm cm}^2/{\rm g}$~\cite{Peter:2012jh}. 

Since superfluids have surface tension, the superfluid core surrounding the brightest central galaxy should be highly isotropic. The source of ellipticity must be the subdominant normal DM component. The normal-normal self-interaction cross section $\sim 0.1~{\rm cm}^2/{\rm g}$ is consistent with the observational bound~\cite{Peter:2012jh}. However, since the normal component only makes up a small fraction of the total DM mass in the central region of galaxy clusters, the rate of self-interaction is considerably smaller, and much larger cross sections are therefore allowed. This clearly deserves further study. Interestingly, the ellipticity has been observed to decrease towards the center of clusters ($r\;\lsim\; 16$~kpc)~\cite{porter1991}, consistent with a highly spherical superfluid core.

\end{itemize}

\section{Discussion}

In this paper we proposed a novel theory of DM superfluidity that reconciles the stunning success of MOND on galactic scales with the triumph of the $\Lambda$CDM model on cosmological scales.
The DM component consists of self-interacting axion-like particles which are generated out-of-equilibrium and remain decoupled from baryons throughout the history of the universe. 
Provided that its mass is sufficiently light and its self-interactions sufficiently strong, the DM can thermalize and form a superfluid in galaxies, with critical temperature of order $\sim$mK. 
The superfluid phonon excitations are assumed to be described by a MOND-like action and mediate a MONDian acceleration on baryonic matter. Superfluidity only occurs at sufficiently low temperature, or equivalently within sufficiently low-mass objects. This naturally distinguishes between galaxies (where MOND is successful) and galaxy clusters (where MOND is not): due to the larger velocity dispersion in clusters, DM has a higher temperature and hence is either in a mixture of superfluid and normal phase, or fully in the normal phase. 

The superfluid interpretation makes the well-known non-analytic nature of the MOND scalar action much more natural. The phonons of the unitary Fermi gas,
which has attracted much excitement in the cold atom community recently~\cite{UFGreview}, are also governed by a non-analytic kinetic term (with $5/2$ power
instead of $3/2$ for our DM superfluid). The DM condensate equation of state $P\sim \rho^3$ suggests that our superfluid arises from three-body interactions. It would be fascinating to find precise cold atom systems with the same equation of state as our DM condensate. Practically this would yield important insights on the microphysical interactions that give rise to this particular superfluid. Tantalizingly, it might allow laboratory simulations of the properties and dynamics of galaxies.

The rich physics of superfluidity leads to a number of observational signatures that can potentially distinguish our scenario from ordinary MOND and/or standard $\Lambda$CDM:
numerous low-density vortices in galaxies; merger dynamics depending on the infall velocity vs phonon sound speed; distinct mass peaks in bullet-like cluster mergers, corresponding to superfluid and normal components; interference patters in super-critical mergers. Studying these observables with numerical simulations promises to be fascinating.\\

\noindent {\bf Acknowledgements:} We thank Niayesh Afshordi, Asimina Arvanitaki, Luc Blanchet, Adrienne Erickcek, Henry Glyde, Paul Hamilton, Philipp Haslinger, Lam Hui, Matthew Jaffe, Bhuvnesh Jain, Arthur Kosowsky, Werner Krauth, Michele Maggiore, Stacy McGaugh, Alberto Nicolis, Marcel Pawlowski, James Peebles, Ravi Sheth, David Spergel, Paul Steinhardt and Matias Zaldarriaga. We are particularly grateful to Benoit Famaey for stimulating discussions on observations, and to Randy Kamien and Tom Lubensky for patiently teaching us about soft condensed matter theory. J.K. is supported in part by NSF CAREER Award PHY-1145525 and NASA ATP grant NNX11AI95G. L.B. is supported by funds provided by the University of Pennsylvania.

\section*{Appendix: Why $\Phi^6$ Fails to Give MOND}
\renewcommand{\theequation}{A-\Roman{equation}}
\setcounter{equation}{0} 

In this Appendix we show that a complex scalar field with hexic interactions yields a phonon action with the
desired $3/2$ power but with the wrong sign to give the MOND phenomenon. Our starting point is the relativistic action
\be
\mathcal{L}=-|\partial_\mu \Phi |^2-m^2 |\Phi |^2-\frac{\lambda}{3}|\Phi|^6\,,
\ee
This theory is invariant under global $U(1)$ symmetry, with associated conserved charge being the number of particles.
Making the replacement $\Psi = \Phi e^{imt}$ and taking the non-relativistic limit, the theory becomes
\be
\mathcal{L}=\frac{i}{2}\left(\Psi\partial_t \Psi^*-\Psi^* \partial_t \Psi\right)-\frac{|\vec{\nabla} \Psi |^2}{2m}-\frac{\lambda}{24m^3}|\Psi |^6\,.
\label{NRtheory}
\ee
The equation of motion is a nonlinear Schr\"odinger's equation,
\be
-i\partial_t \Psi+\frac{\vec{\nabla}^2 \Psi}{m}-\frac{\lambda}{8m^3}|\Psi |^4\Psi =0\,.
\ee
This equation possesses the following homogeneous background solution which describes the BEC at zero temperature
\be
\Psi_0=\sqrt{2m}ve^{i\mu t}\,, 
\label{BECsolution}
\ee
where $\mu\equiv \frac{\lambda v^4}{2m}$ is the chemical potential. Meanwhile $v$ is related to the number density of particles in the condensate, $n=2mv^2$, which in turn is the Noether charge density of the spontaneously broken $U(1)$ symmetry. 

To study the spectrum of perturbations around~\eqref{BECsolution}, we can expand as follows
\be
\Psi=\sqrt{2m}(v+\rho)e^{i(\mu t+\phi)}\,,
\label{pert}
\ee
where $\rho$ is the perturbation of the order parameter, while $\phi$ is the Goldstone boson.\footnote{Strictly speaking,~\eqref{BECsolution} spontaneously breaks the diagonal combination of the internal $U(1)$ and time translation. Therefore $\phi$ is the Goldstone boson of this symmetry.} 
Substituting into \eqref{NRtheory} we obtain
\be
\mathcal{L}=-(\vec{\nabla}\rho)^2+2m(v+\rho)^2\left[ \mu+\dot{\phi}-\frac{(\vec{\nabla} \phi)^2}{2m} \right]-\frac{\lambda}{3}(v+\rho)^6\,.
\label{NRptheory}
\ee
The low energy spectrum of the theory can be deduced as usual by linearizing the equations of motion and computing the characteristic determinant. The analysis shows there is one dynamical degree of freedom in the spectrum, with dispersion relation\footnote{In a fully relativistic treatment, one would find an additional degree of freedom with mass $2m$. This mode is of course too heavy to be captured by the non-relativistic treatment.} 
\be
\omega^2=\frac{\lambda v^4}{m^2}k^2+\mathcal{O}(k^4)\,.
\label{disp}
\ee
Thus $\lambda>0$ is necessary for stability.

The effective theory of the Goldstone can be obtained by integrating out $\rho$. To leading order in the derivative expansion the $(\vec{\nabla}\rho)^2$ term can be ignored, with resulting action
\be
\mathcal{L}=\frac{4}{3}m\left( \mu+\dot{\phi}-\frac{(\vec{\nabla}\phi)^2}{2m} \right)\left( \frac{2m}{\lambda} \left[ \mu+\dot{\phi}-\frac{(\vec{\nabla}\phi)^2}{2m} \right] \right)^{1/2}\,.
\label{almostMOND}
\ee
As a consistency check let us linearize the theory and compare the result to the dispersion relation~\eqref{disp}. The quadratic Lagrangian for $\phi$ reduces to
\be
\mathcal{L}_{\rm quad}=m \left( \frac{2m\mu}{\lambda} \right)^{1/2}\left( \frac{1}{2\mu}\dot{\phi}^2-\frac{1}{m}(\vec{\nabla}\phi)^2 \right)\,.
\label{L2}
\ee
Perturbations are stable for $\mu>0$, which is guaranteed by $\lambda>0$. Taking into account the explicit expression~\eqref{BECsolution} for the chemical potential we recover the dispersion relation~\eqref{disp}.

Notice that~\eqref{almostMOND} looks very similar to~\eqref{Lphon}. It involves the correct fractional power needed for the MOND action. Unfortunately, 
because of the requirement $\lambda> 0$ the gradient term can never dominate over $\mu$, and the would-be MOND regime is inaccessible.
One may be tempted to focus on $\lambda<0$ instead, since in that case the limit of large gradients appears to be well defined. Moreover, we even obtain the correct equation of state for the condensate when we set $\phi=0$, taking into account that $\mu/\lambda>0$. However, according to~\eqref{L2} the perturbations around the condensate have a ghost-like kinetic term for $\mu<0$. The physical origin for this instability is very simple --- $\lambda < 0$ corresponds to an attractive interaction between bosons, hence the homogeneous BEC is unstable against collapse. 

In contrast the theory with $|\partial\Phi|^6$ interactions studied in Sec.~\ref{phi6} precisely gives the phonon theory~\eqref{Lphon} and has stable perturbations around the homogeneous BEC background.

\end{document}